\documentclass[%
 reprint,
 amsmath,amssymb,
 aps,
floatfix,
]{revtex4-2}
\pdfoutput=1
\usepackage{amsmath}
\usepackage{graphicx}
\usepackage{dcolumn}
\usepackage{bm}
\usepackage{hyperref}
\hypersetup{colorlinks,allcolors=black}
\usepackage{comment}
\usepackage{subcaption}
\usepackage{float}

\begin{document}
\preprint{APS/123-QED}

\title{Optimization of a Lossy Microring Resonator System for the Generation of Quadrature-Squeezed States}

\author{Colin Vendromin}
\email{colin.vendromin@queensu.ca}
\author{Marc M. Dignam}
\affiliation{%
Department of Physics, Engineering Physics, and Astronomy, Queen's University, Kingston, Ontario K7L 3N6, Canada
}%

\date{\today}

\begin{abstract}
The intensity buildup of light inside a lossy microring resonator can be used to enhance the generation of squeezed states via spontaneous parametric downconversion (SPDC). In this work, we model the generation of squeezed light in a microring resonator that is pumped with a Gaussian pulse via a side-coupled channel waveguide. We theoretically determine the optimum pump pulse duration and ring-to-channel coupling constant to minimize the quadrature noise (maximize the squeezing) in the ring for a fixed input pump energy. We derive approximate analytic expressions for  the optimal coupling and pump pulse duration as a function of scattering loss in the ring. These results will enable researchers  to easily determine the optimal design of microring resonator systems for the generation of quadrature-squeezed states. 
\end{abstract}

\maketitle

\section{\label{intro}Introduction}
Squeezed states are a type of nonclassical light that are characterized by squeezing of the quantum uncertainty in a given quadrature below the level of vacuum noise. They can be used in a variety of contexts, including in applications where quadrature noise is a major concern, such as optical communications \cite{Slavik2010All-opticalSystems} and interferometers \cite{Tan2014EnhancedStates,schnabelSqueezeStates, Aasi2013EnhancedLight}. Squeezed states can also be used as the starting point to create entangled states of light. Weakly-squeezed states can be used as a source of entangled photons, which can be used for quantum teleportation \cite{Qteleportation} and quantum cryptography \cite{Qcrypto}. Single-mode squeezed states can be combined using waveguide couplers to create quadrature-entangled states \cite{Masada2015Continuous-variableChip}. In addition, two-mode quadrature-squeezed states are a source of continuous variable (CV) entanglement, which can also be used for quantum computation \cite{quantumcomputingsqueezelimit} and quantum information \cite{BraunsteinQuantumInformationCV}; such states are important as they are generally more robust to loss than two-photon entangled states \cite{Braunstein1998TeleportationVariables}. 

One way to generate squeezed states of light is via spontaneous parametric down conversion (SPDC), where a strong coherent pump field interacts with a material that has a $\chi^{(2)}$ nonlinearity \cite{squeezedstatesviaSPDC}. The conversion efficiency of pump photons into signal and idler pairs can be enhanced by enclosing the nonlinear interaction within a cavity that is resonant with the pump. In this case, if it is a multimode cavity, where a second mode is resonant at the signal and idler frequencies, then it can play a dual role, by ensuring that essentially all generated pairs end up in a single cavity mode. 

Ring resonators side-coupled to a waveguide have been shown to enhance spontaneous parametric down conversion efficiency \cite{Yang2007EnhancedResonator}. Thus, they are promising structures for on-chip applications such as entangled photon pair generation for quantum communication \cite{Lu2019Chip-integratedCommunication} and generating squeezed light for discrete and CV entanglement \cite{Vernon2018ScalableSampling, VaidyaBroadbandDevice, Samara2019High-RateMicroresonators, Guo2016ParametricChip}. The schematic diagram of a side-coupled ring resonator is shown in Fig. \ref{fig:ringresonator}. The ring waveguide  has a radius chosen such that it has resonant modes at the frequencies of the pump and the squeezed light. The straight waveguide (channel) and ring are in proximity to each other, such that pump and squeezed light can be evanescently coupled in and out of the resonator.

Considerable theoretical work has been done on a Hamiltonian treatment of SPDC and spontaneous four-wave mixing in lossy microring resonators \cite{Yang2007GeneratingResonators, Vernon2015SpontaneousResonators,Vernon2015StronglyResonators,alsingEntangleandSqueezeLossyRing}. The general approach is to solve the Heisenberg equations of motion for the mode operators in the ring and channel. This procedure is applicable to both the weak pumping limit for generating entangled photon pairs and  the strong pumping limit for generating quadrature squeezing. For example  single-mode quadrature squeezing of -10dB  in the channel of a lossy SiN microring resonator was recently shown to be theoretically achievable \cite{Vernon2018ScalableSampling}, using a 100pJ Gaussian input pulse of duration 30ps. Experimentally, about 4dB \cite{VaidyaBroadbandDevice} to 5dB \cite{Dutt2015On-ChipSqueezing} of squeezing has been inferred on-chip with SiN microring resonators.  Both the theory and experimental demonstration of quadrature squeezing in lossy microring resonators provides a promising path forward for creating a practical CV entangled states for quantum computing applications. 

Recent experimental work has demonstrated that one can tune the squeezing level generated in coupled ring resonators; by increasing the coupling efficiency, Dutt \textit{et al.} \cite{DuttRingSqueezingVscoupling} demonstrated experimentally an increase of the on-chip squeezing level in a SiN resonator from $-0.9$dB to $-3.9$dB.
Although this and other work demonstrate the promise of ring resonators for generating squeezed light, it appears that very little has been done on the \textit{optimization} of the ring resonator system to obtain maximum squeezing. 

In this paper, we theoretically study the quadrature squeezing inside a lossy ring resonator pumped by a Gaussian input pulse. We focus on the optimization of the pump pulse duration and ring-channel coupling, in order to achieve the conditions that maximize the squeezing in the presence of scattering loss. 

We consider the case of squeezed-state generation via SPDC in a single mode of the ring. To allow us to compare the squeezing achieved for different pump durations, in all that follows, the energy of the input pulse is held constant when the pulse duration is changed. We model the dynamics of the density operator for the state in the ring in the presence of loss using the Lindblad master equation for a cavity with a single lossy mode.  It has recently been shown that the general solution to this Lindblad master equation is a single-mode squeezed thermal state \cite{Seifoory2017SqueezedCavities} characterized by a time-dependent squeezing amplitude, squeezing phase, and a thermal photon number. Using this solution, we model the squeezed thermal state in the ring resonator as a function of time, and derive an approximate analytic expression for the maximum squeezing in the presence of loss.

Our theoretical approach is somewhat different from what is commonly done in the literature. The strength of our method is that, because we know that the density operator inside the ring is always a squeezed thermal state, the time-dependent properties of the state in the ring, such as the variance of the quadrature operator and expectation value of the number operator, can be easily determined by simply solving for the time dependence of the thermal photon number and squeezing parameter of the state. Of course, our study is restricted to a single-mode squeezed state in the ring, but this condition is easily satisfied by limiting the bandwidth of the input pulse, and carefully phase-matching the desired pump mode and squeezed light mode in the ring. 

Using our exact solution for the time evolution of the state, we derive approximate but accurate analytic expressions for the optimum coupling value and optimum pump pulse duration for a fixed pump energy. We show that they are in excellent agreement with full numerical simulations when the pump and ring configuration is relatively close to the optimal. We find that the optimum pulse duration depends on the loss in the ring and is in the range of of $10$ to $60$ times the ring round-trip time. We also show that the optimum coupling is slightly below critical coupling (undercoupling).

The paper is organized as follows. In section \ref{sec:ringtheory} we review the theory of the coupling of a pulsed classical pump field from a channel waveguide into a ring resonator, discuss practical limitations on the pump pulse duration for generation in a single-mode, and determine the exact and approximate expressions for the time-dependent pump field inside the lossy ring. In section \ref{sec:squeezingtheory} we present the theory behind the generation of a squeezed thermal state in a single leaky mode for a pulsed pump. In section \ref{sec:results} we model the system and develop approximate analytic expressions for the optimal pulse duration, coupling constant and quadrature noise for a given ring loss. Finally, in section \ref{sec:conclusions} we present our conclusions.
\section{\label{sec:theory}Theory}
In this section we present the theory behind the generation of squeezed light inside a ring resonator. The system consists of a ring resonator waveguide of radius $R$ side-coupled to a straight waveguide (the channel) (see Fig. \ref{fig:ringresonator}).  Both waveguides are made from a material with a nonlinear $\chi^{(2)}$ response. We treat the ring resonator as an optical cavity that generates squeezed light in a single leaky mode. The mode is leaky due both to scattering loss and coupling to the channel. The input field to the system is a classical pump pulse  ($E_1(t)$) propagating in the channel. The bandwidth of the input pulse is limited such that it only couples appreciably into a single mode inside the ring, with frequency, $\omega_P$. Once inside the ring, the pump will produce squeezed light in a separate mode with frequency, $\omega_S$, that is half the frequency of the pump, \textit{i.e.} $\omega_S = \omega_P/2$. In section \ref{sec:ringtheory} we study the frequency response of the ring using a transfer matrix approach in the presence of loss, and derive exact and approximate expressions for the time-dependent pump field inside the ring. In section \ref{sec:squeezingtheory} we give the solution to the Lindblad master equation for the quantum state of light generated inside the ring.

\subsection{Time-Dependent Pump Field Inside the Ring Resonator}
\label{sec:ringtheory}
In this section we present the theory to obtain the time dependence of the pump field inside the ring resonator and examine the dependence of the field build-up in the ring on the pump pulse duration, the scattering loss in the ring, and the coupling between the channel and ring waveguides.

The classical pump pulse field, $E_1(t)$, incident on the ring resonator is taken to be a classical Gaussian pulse of
the form
\begin{equation}
   E_1(t) =E_1^{(+)}(t) + E_1^{(-)}(t),\nonumber
\end{equation}
where
\begin{equation}
 \label{eq:incidentpulse}
   E_1^{(+)}(t)=E_0\sqrt{\frac{T_R}{\tau}} \exp\left(-2\ln\left(2\right)\frac{t^2}{\tau^2}\right)\exp(-i\omega_P t),
\end{equation}
and $E^{(-)}(t) =\left[ E^{(+)}(t)\right]^*$. Here $\tau$ is the duration of the pulse (FWHM of the intensity), $\omega_P$ is the pulse carrier frequency, $T_R$ is the ring round-trip time (discussed in more detail below), and $E_0$ is the amplitude of the pulse. The factor of $1/\sqrt{\tau}$ is included so that the energy of the pulse is independent of the pulse duration. We do this so that we can study the squeezing level in the ring for many different pumping durations, with a constant amount of energy going into the system. In the following, only the positive frequency part of the input field is needed, because we are using the rotating wave approximation. 

In calculating the coupling of the field in and out of the ring, it is easier to work in the frequency domain. We define the Fourier transform of the time-dependent field as
\begin{eqnarray}
    \label{eq:FT}
    \tilde{E}(\omega) &=& \int_{-\infty}^{\infty}E^{(+)}(t)\exp(i\omega t) dt,
\end{eqnarray}
and the inverse Fourier transform as
\begin{eqnarray}
    \label{eq:INVFT}
    E^{(+)}(t) &=& \frac{1}{2\pi}\int_{-\infty}^{\infty}\tilde{E}(\omega)\exp(-i\omega t) d\omega.
\end{eqnarray}
The Fourier transform of the input pulse of Eq. \eqref{eq:incidentpulse} is
\begin{eqnarray}
    \label{eq:e1w}
    \tilde{E}_1(\omega) = \tilde{E}_0 \sqrt{\frac{\tau}{T_R}} \exp\left(-\frac{(\omega-\omega_P)^2\tau^2}{8\ln2}\right),
\end{eqnarray}
where $\tilde{E}_0\equiv E_0 T_R\sqrt{\pi/(2\ln\left(2\right))}$. The bandwidth $\Delta \omega$ (FWHM in frequency) of the input pulse  is related to the pulse duration $\tau$ by $\Delta \omega = 4\ln{\left(2\right)} / \tau$. 

\begin{figure}[htbp]
    \centering
    \includegraphics[scale = 0.333]{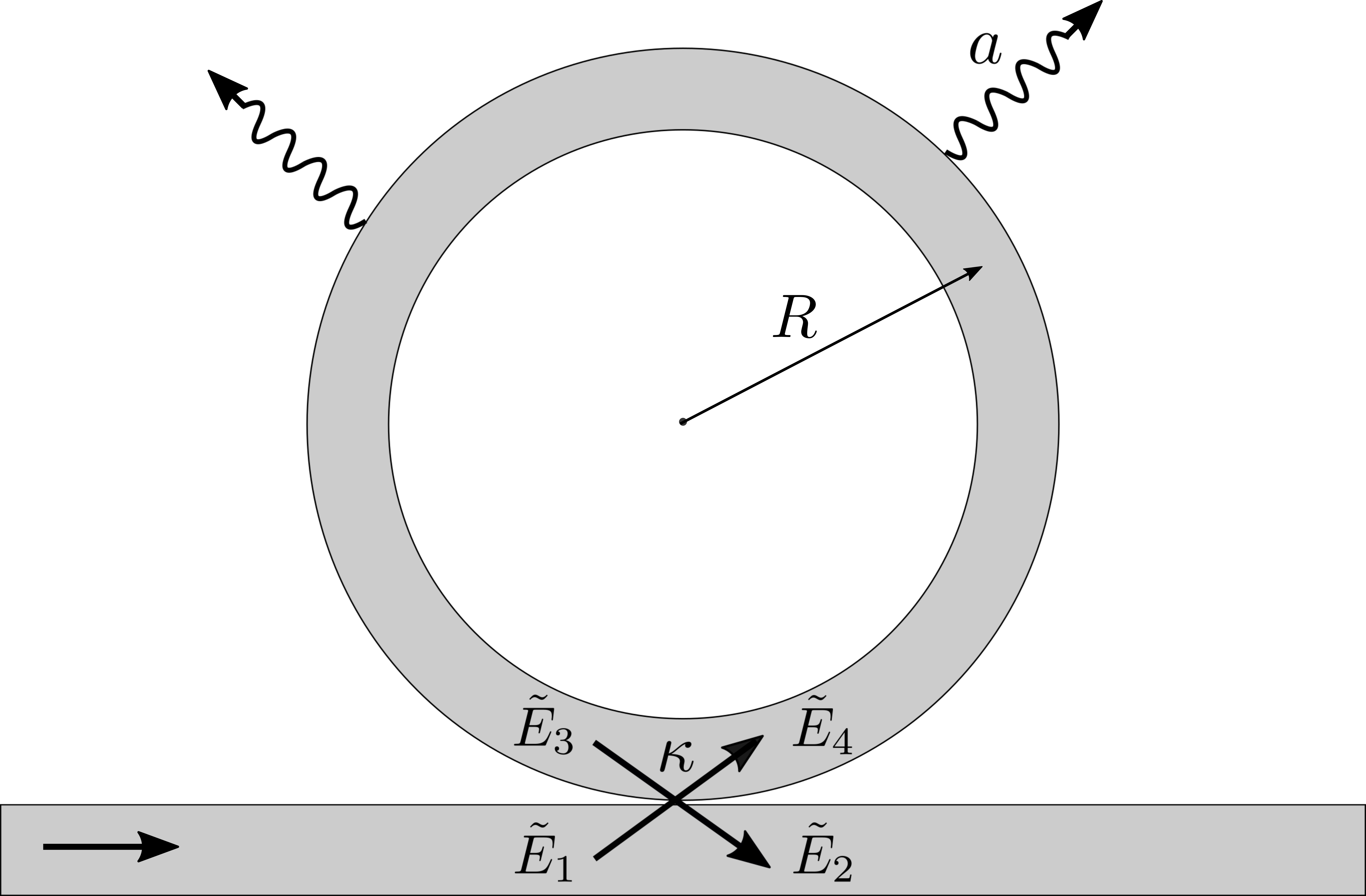}
    \caption{\small Schematic of the ring resonator coupled to a channel waveguide. The field components incident to the coupling point are $\tilde{E}_1$ in the channel and $\tilde{E}_3$ in the ring. The field components leaving the coupling point are $\tilde{E}_2$ in the channel and $\tilde{E}_4$ in the ring. The cross-coupling coefficient is $\kappa$, and the attenuation in the ring is $a$. }
    \label{fig:ringresonator}
\end{figure}

The fields in the ring and channel are assumed to couple at a point, as indicated in Fig. \ref{fig:ringresonator}. The fields incident on the coupling point are $\tilde{E}_1(\omega)$ in the channel and $\tilde{E}_3(\omega)$ in the ring. The fields leaving the coupling point are $\tilde{E}_2(\omega)$ in the channel and $\tilde{E}_4(\omega)$ in the ring. The input and output field components are defined at locations just to the left and right of the coupling point, respectively. The input and output fields are related by a transfer matrix as 
\begin{equation}
    \label{eq:scatteringmatrix}
    \begin{pmatrix}
    \tilde{E}_4(\omega)  \\  \tilde{E}_2(\omega) 
    \end{pmatrix}
    =
     \begin{pmatrix}
    \sigma & i\kappa \\ i\kappa & \sigma
    \end{pmatrix}
     \begin{pmatrix}
     \tilde{E}_3(\omega)  \\  \tilde{E}_1(\omega)
    \end{pmatrix},
\end{equation}
where $\sigma$ and $\kappa$ are real numbers called the self- and cross- coupling coefficients, respectively. This is the form of the transfer matrix that is commonly used \cite{Heebner2004DistributedOptics}. The coupling is assumed to occur at a single point, so the field components that pass through the coupling point and stay in the same waveguide do not acquire a phase. However, the field components that cross-over into the other waveguide at the coupling point do acquire the phase $i$. This phase is needed in order to conserve power across the coupling point ($i.e.$ the transfer matrix must be unitary).
 Additionally, the coupling is assumed to be lossless, so we obtain the relation $|\sigma|^2 + |\kappa|^2 = 1$. The fields $\tilde{E}_4(\omega)$ and $\tilde{E}_3(\omega)$ are related by,
\begin{eqnarray}
    \label{eq:e3w}
    \tilde{E}_3(\omega) = a \exp(i\Theta)\tilde{E}_4(\omega).
\end{eqnarray}
Here, $a$, is the field attenuation after one circuit of the ring (excluding any coupling to the straight waveguide); this is related to the scattering power-loss coefficient, $\alpha_{\rm sc}$, in the ring by $a = \exp(-\alpha_{\text{sc}}2\pi R/2)$. In what follows, we assume that $a$ is frequency independent, and also that $a$ and $\kappa$ are independent of each other. The single-circuit phase shift $\Theta$ in the ring is given by $\Theta =2\pi Rk$, where $k = 2\pi n_{\rm eff}/\lambda$, where $n_{\rm eff}$ is the effective index of refraction for the pump mode in the ring and $\lambda$ is the free space wavelength. The phase shift can also be expressed as,
\begin{equation}
    \label{eq:thephase}
    \Theta= \omega T_R, 
\end{equation}
where $T_R = n_{\rm eff}2\pi R /c$ is the ring round-trip time. For light that is on resonance with a mode in the ring, the phase shift is $\Theta = 2\pi m$, where $m$ is a positive integer (the mode number). Thus, in order to ensure that the pump frequency is on resonance with the ring, it is chosen to be $\omega_P = 2\pi m_P /T_R$, where $m_P$ is the pump mode number. In all that follows, we will scale the time, the pump duration, and the pump pulse amplitude by the round-trip time $T_R$; consequently, all of the results that follow are independent of the ring radius and mode number. 

We choose the frequency of the signal and idler photons to both be $\omega_S =\omega_P/2$ (where $S$ stands for ``squeezed light"), such that the mode number for the squeezed light is $m_S = m_P/2$. The coupling coefficients are assumed to be frequency independent. This is a good approximation as long as the pump pulse is in a single mode. We assume that the ring waveguide dimensions have been chosen such that the squeezed light mode has the same $n_{\rm eff}$ as the pump mode (\textit{i.e.}, they are phase matched). This has been demonstrated in an AlN ring resonator \cite{Guo2016ParametricChip} for a waveguide with a height of $1{\rm \mu m}$ and a width of $1.10 {\rm \mu m}$, and in AlGaAs nanowaveguides \cite{Rutkowska2011SecondNanowaveguides}.

Using Eqs. \eqref{eq:scatteringmatrix} and \eqref{eq:e3w}, we find that the field inside the ring is given by
\begin{equation}
    \label{eq:fieldinring}
    \tilde{E}_3(\omega)= \frac{i\sqrt{1-\sigma^2}\, a \exp(i\omega T_R)}{1-\sigma a \exp(i \omega T_R)}\tilde{E}_1(\omega).
\end{equation}
The ratio of intensity inside the ring to the incident intensity in the channel is defined as the buildup factor, 
\begin{equation}
    \label{eq:buildupfactor}
    \mathcal{B}(\omega) \equiv \left|\frac{\tilde{E}_3(\omega)}{\tilde{E}_1(\omega)}\right|^2 = \frac{(1-\sigma^2)a^2}{1-2\sigma a \cos(\omega T_R) + \sigma^2 a^2}.
\end{equation}
It is maximized for light that is on resonance with the ring, \textit{i.e.} $\cos(\omega T_R) = 1$. Using $\omega=\omega_P$ in Eq. \eqref{eq:buildupfactor} gives the maximum value of the buildup factor,
\begin{equation}
    \label{eq:idealbuildupfactor}
    \mathcal{B}(\omega_P) =  \frac{\left(1-\sigma^2\right)a^2}{\left(1-\sigma a\right)^2}.
\end{equation}
The value of $a$ that maximizes Eq. \eqref{eq:idealbuildupfactor} is $a = \sigma$. This is known as \textit{critical coupling}.

\begin{figure}[htbp]
        \centering
            \includegraphics[width=0.48\textwidth]{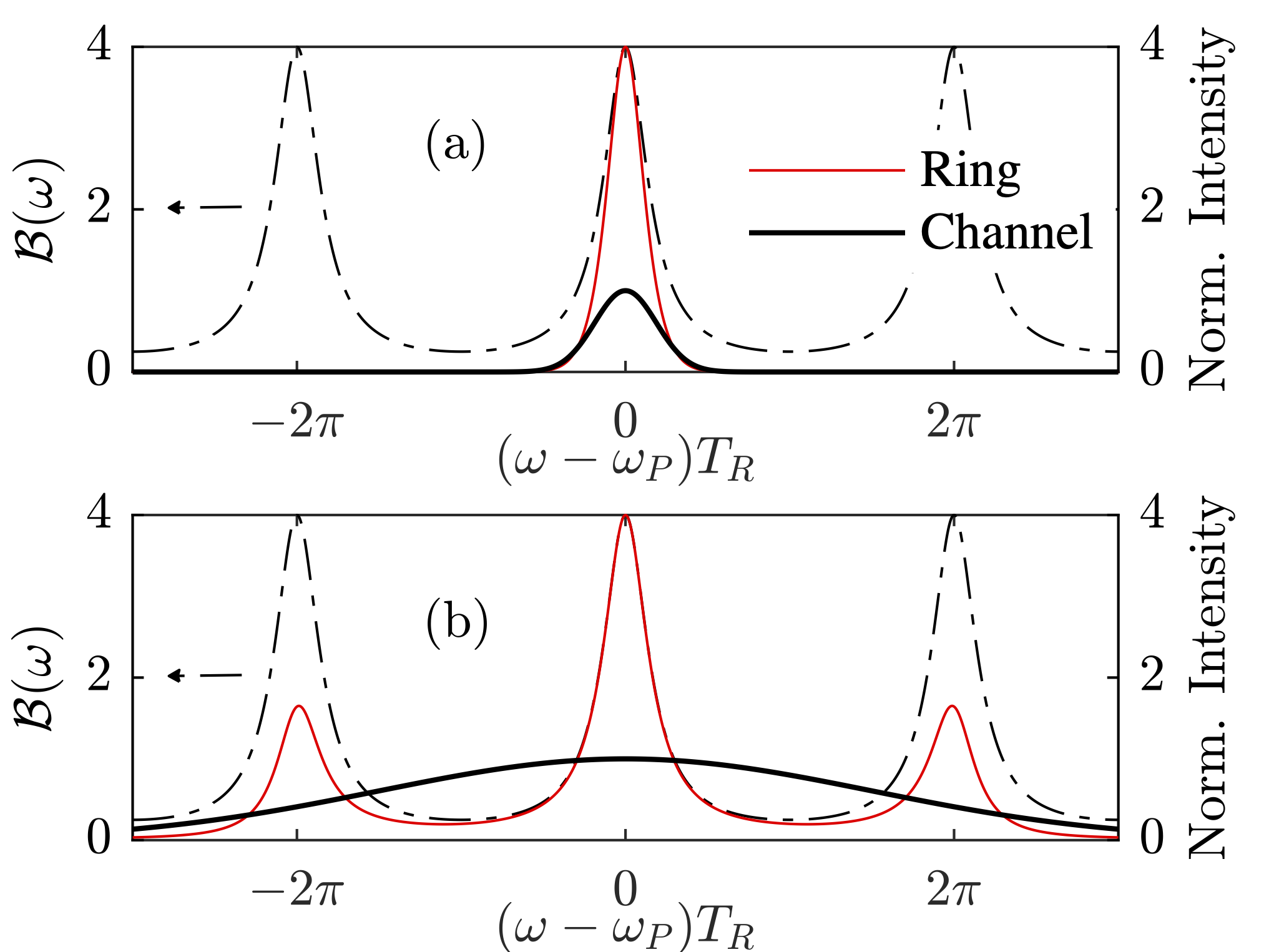}
       \caption{\small The intensity buildup (in three ring modes) of the pump pulse with a duration of (a) $\tau = 2 T_R$ and (b) $\tau = T_R/4$. The normalized intensity of the input pulse in the channel is $|\tilde{E}_1|^2/\tilde{E}^2_0$ (thick curve), and the normalized intensity of the pulse in the ring, after its intensity has built up, is $|\tilde{E}_3|^2/\tilde{E}^2_0$  (thin red curve). The buildup factor (dashed line) and intensity are calculated with $\sigma = 0.6$ and $a=1$.}
\label{fig:buildup}
\end{figure}

To ensure that the squeezed light will be generated mostly in a single mode with frequency $\omega_S$ we require that the pump pulse almost exclusively couples into a single mode in the ring with frequency $\omega_P$. In Fig. \ref{fig:buildup}(a) we demonstrate that with an incident pulse with duration $\tau = 2 T_R$ (thick line), virtually all of the pulse intensity couples into a single ring resonance (thin red line). In contrast, in Fig. \ref{fig:buildup}(b) we show that by reducing the pulse duration to $\tau = T_R/4$, the broadening of the pulse in frequency causes some of its intensity to couple into adjacent modes. Thus, in all that follows, we restrict ourselves to pulses with duration $\tau \ge T_R$ to ensure  the squeezed light is generated almost entirely in a single mode. Although two-mode squeezed light could also be generated in a number of different mode pairs that satisfy energy conservation, we assume that generation in those other modes is suppressed because they are not well phase matched. 

The intensity decay rate $\Gamma$ of  light in the ring cavity is given by,
\begin{equation}
\label{eq:decayrate}
\Gamma \equiv\frac{\alpha_{\text{tot}}2\pi R}{T_R} =\frac{1}{T_R} \left[\ln\left(\frac{1}{\sigma^2}\right) + \ln\left(\frac{1}{a^2}\right)\right],
\end{equation}
where $\alpha_{\rm tot}$ is the total loss coefficient for the ring. It is given by $\alpha_{\rm tot} = \alpha_{\rm {sc}} + \alpha_{\rm {cpl}}$, where $\alpha_{\rm sc}$ is given above, and $\alpha_{\rm cpl}$ is defined by the equation $\sigma = \exp\left(-\alpha_{\rm cpl}2\pi R/2\right)$ \cite{2008OpticalMicroresonators} and is the power-loss coefficient due to light coupling out of the ring into the channel. To obtain strong squeezing in the ring, the intensity decay rate multiplied by the round-trip time must be small, \textit{i.e.}, $\Gamma T_R \ll 1$. If the loss is small enough such that $(1 - \sigma a) \ll 1$ then from Eq. \eqref{eq:decayrate}, we obtain,
\begin{equation}
    \label{eq:approxdecayrate}
  \Gamma  \approx\frac{2(1-\sigma a)}{T_R}.
\end{equation}
The decay rate $\Gamma$ gives an estimate of the width of the peaks in the the buildup factor.

The time-dependent pump field, $E_3(t)$, inside the ring just to the left of the coupling point (see Fig. \ref{fig:ringresonator}) is calculated by taking the inverse Fourier transform of Eq. \eqref{eq:fieldinring}, giving:
\begin{eqnarray}
    \label{eq:e3time}
    E_3\left(\tilde{t}\,\right) &=&  \frac{i\kappa aE_0}{\sqrt{\pi}}\exp\left(-i2\pi m_P \tilde{t}\,\right)\sqrt{\frac{\tilde{\tau}}{8\ln2}}  \times \nonumber 
    \\
    &\times& \int_{-\infty}^{\infty} d\Omega\frac{\exp\left(-\Omega^2\tilde{\tau}^2/(8\ln2) - i\Omega \tilde{t}\,\right)}{\exp(-i\Omega) - \sigma a},
\end{eqnarray}
where $\Omega \equiv (\omega-\omega_P)T_R$, $\tilde{t} \equiv t/T_R$, and $\tilde{\tau}\equiv \tau/T_R$. The integral is real because we integrate $\Omega$ from $-\infty$ to $\infty$. This is the general expression that we use in our simulations. In the low-loss limit, where $(1-\sigma a)\ll 1$,  the integral in Eq. \eqref{eq:e3time} can be evaluated using Voigt functions \cite{analyticringpulse} (see Appendix \ref{ringpulsederiv}), and we obtain the approximate expression
\begin{eqnarray}
\label{e3timeapprox}
   |E_3\left(\tilde{t}\,\right)| =  \frac{\sqrt{\pi}\kappa a \tilde{\tau}\,{\rm e}^{z\left(\tilde{t}\,\right)^2}{\rm erfc}\left[\,z\left(\tilde{t}\,\right)\right]}{\sqrt{8\ln2}}\big|E_1^{(+)}\left(\tilde{t}\,\right)\big|, 
\end{eqnarray}
where
\begin{equation}
\label{ztransformation}
    z\left(\tilde{t}\,\right)\equiv \frac{(1-\sigma a)\tilde{\tau}}{\sqrt{8\ln(2)}}-\frac{\sqrt{8\ln(2)}\tilde{t}}{2\tilde{\tau}},
\end{equation}
and ${\rm erfc}\,\left(x\right) = 1 - {\rm erf} \,\left(x\right)$, where ${\rm erf}\,\left(x\right)$ is the error function. In the following sections, we shall use this expression to optimize the incident pump pulse duration to achieve the greatest nonlinear response in the ring. We note parenthetically that Eq. \eqref{e3timeapprox} would also be useful for calculating classical nonlinear processes such as second harmonic generation or parametric downconversion in a ring resonator, using the undepleted pump approximation.

In this section, we have derived an expression for the time-dependent pump field inside the ring, which we shall use in the following section to calculate the generation of the squeezed state. 

\subsection{Quadrature Squeezing Inside a Lossy Ring Cavity }
\label{sec:squeezingtheory}
In this section we present the main theory behind quadrature squeezing inside the ring.

The Hamiltonian for light inside the ring, using the undepleted pump approximation, is given by \cite{quantumopticsGarrison}
\begin{eqnarray}
    \hat{H} &=& \hat{H}_0 + \gamma E_3(t) \hat{b}^{\dagger 2} + \gamma^*E^*_3(t){\hat{b}}^2, \label{eq:Hnoloss}
\end{eqnarray}
where the interaction-free part of the Hamiltonian is $\hat{H}_0= \hbar \omega_S \hat{b}^\dagger \hat{b}$, and the last two terms account for the SPDC process. The  operator $\hat{b}$ is the annihilation operator for the squeezed light photons in the ring. The nonlinear coupling coefficient between the pump, $E_3(t)$, and squeezed light is $\gamma = \hbar \omega_P \chi^{(2)}_{\text{eff}}/n^2_{\rm eff}$, where $\chi^{(2)}_{\text{eff}}$ is an effective nonlinear susceptibility that depends on the intrinsic nonlinear susceptibility of the ring material and spatial mode profiles in the ring \cite{Seifoory2019CounterpropagatingWaveguides}. Note that we neglect any nonlinear interactions in the channel waveguide, because the pump intensity is much smaller there. The pump field is given in Eq. \eqref{eq:e3time}, where only the positive frequency part is used, as we are using the rotating wave approximation.  

The effects of scattering and coupling losses on the dynamics of the generated light in the ring can be modelled using the Lindblad master equation for the density operator $\hat{\rho}$ \cite{openQsystemsBreuer}:
\begin{equation}
    \label{eq:lindblad}
    \frac{d\hat{\rho}}{dt}=-\frac{i}{\hbar}\left[\hat{H},\hat{\rho}\right] + \Gamma\left( \hat{b}\hat{\rho}\hat{b}^\dagger -\frac{1}{2} \hat{b}^\dagger\hat{b}\hat{\rho}-\frac{1}{2} \hat{\rho}\hat{b}^\dagger\hat{b}\right),
\end{equation}
where $\Gamma$ is the decay rate for the squeezed light generated in the cavity. It is given in Eq. \eqref{eq:decayrate}, where now $\sigma$ and $a$ correspond to the coupling and loss parameter for the squeezed light. For simplicity, we have assumed that the squeezed light and the pump have the same coupling and loss parameters, but it is straightforward to generalize this within our theory. The effects of thermal photon populations are negligible at room temperature for the optical frequencies of interest, and so they are not included. 

\begin{figure}[htbp]
 \centering
\includegraphics[scale=1]{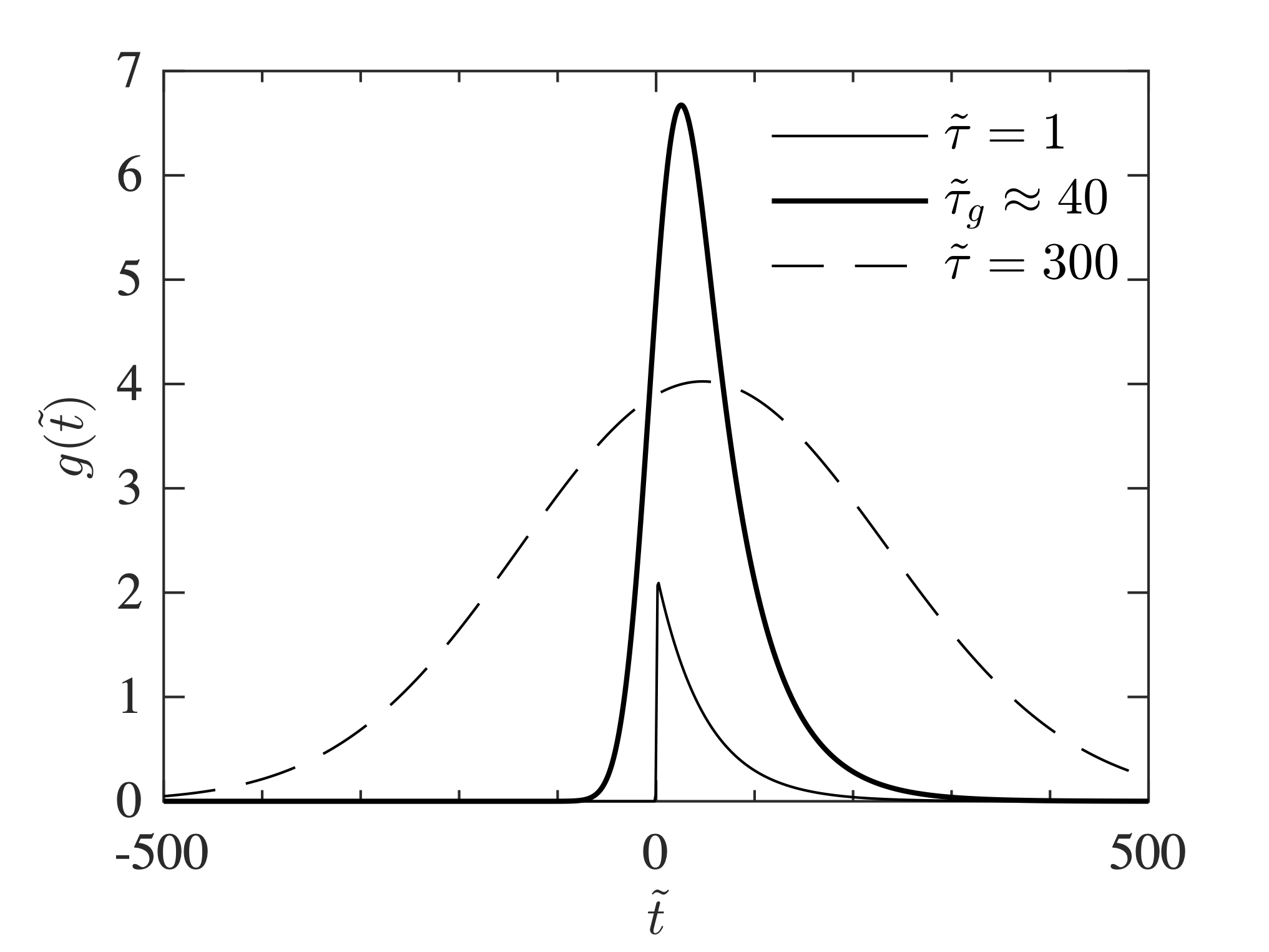} 
\caption{\small The pumping strength $g\left(\tilde{t}\,\right)$ in the ring for $\sigma=a=0.99$ (critical coupling) generated with a short input pulse ($\tilde{\tau} = 1$) (solid thin line), a pulse $\tau =\tau_g$ that gives the highest peak in $g$ (solid bold line), and a long pulse ($\tilde{\tau} =300$) (dashed line). }
\label{fig:e3time}
\end{figure}

It was recently shown \cite{Seifoory2017SqueezedCavities} that the exact solution to Eq. \eqref{eq:lindblad} for the Hamiltonian given in Eq. \eqref{eq:Hnoloss} is a \emph{squeezed thermal state}, which can be written as,
\begin{equation}
    \label{eq:solution}
    \hat{\rho}(t) = \hat{S}(\xi(t))\hat{\rho}_{\text{th}}(\beta(t))\hat{S}^\dagger(\xi(t)),
\end{equation}
where 
\begin{equation}
    \hat{\rho}_{\text{th}}(\beta(t)) = \left(1-{\rm e}^{-\beta(t)\hbar\omega_P/2}\right)^{-1}{\rm e}^{-\beta(t)\hat{H}_0}
\end{equation}
is the density operator for a thermal state at an effective time-dependent temperature $ T(t) = (k_{\rm B}\beta(t))^{-1}$, where $k_{\rm B}$ is the Boltzmann constant. In what follows, rather than use the effective temperature, we characterize this thermal state by the average thermal photon number, which is given by 
\begin{equation}
    \label{eq:nth}
    n_{\text{th}}(t) = \left( {\rm e}^{\beta(t)\hbar\omega_P/2} - 1 \right)^{-1}.
\end{equation}
The operator $\hat{S}$ is a unitary squeezing operator, given by  
\begin{equation}
\label{eq:sqopt}
\hat{S}(\xi(t)) = \exp\frac{1}{2}\left(\xi^*(t)\hat{b}^2 - \xi(t) \hat{b}^{\dagger 2}\right),
\end{equation}
with a complex squeezing parameter $\xi(t) = u(t)\exp(i\phi(t))$. The form of the state given in Eq. \eqref{eq:solution} is only a solution to the Lindblad master equation if the squeezing amplitude $u$, squeezing phase $\phi$, and average thermal photon number $n_{\rm th}$ obey the following three coupled first order differential equations:
\begin{eqnarray}
\label{eq:sqamp}
\frac{1}{\Gamma}\frac{du(t)}{dt}
&=& \frac{g(t)}{2} -\frac{ \cosh u(t) \sinh u(t)}{2n_{\text{th}}(t)+1},
\\
\frac{d\phi(t)}{dt} &=& -\omega_P,  \label{eq:sqphase}
\\
\frac{1}{\Gamma}\frac{dn_{\text{th}}(t)}{dt} &=&\sinh^2u(t) - n_{\text{th}}(t) \label{eq:thmnum}.
\end{eqnarray}
Here,
\begin{equation}
\label{eq:pump}
g(t) \equiv\frac{4|\gamma||E_3(t)|}{\hbar \Gamma}
\end{equation}
 is a dimensionless function of time that we will refer to as the pumping strength \cite{Seifoory2017SqueezedCavities}; it is the  ratio of the pumping rate to the total decay rate of the squeezed light in the cavity. It is constructed such that when $g(t) = 1$,  the rate of signal generation in the ring equals the signal loss out of the ring. Using the approximate expression for the field in Eq. \eqref{e3timeapprox}, we can write the pumping strength as,
 \begin{eqnarray}
     \label{pumpapprox}
     g\left(\tilde{t}\,\right)&=&g_0\frac{\kappa a  }{ \tilde{\Gamma}}\sqrt{\frac{\tilde{\tau}}{8\ln2}}\exp\left(\frac{-2\ln(2)\tilde{t}^2}{\tilde{\tau}^2}\right)\times \nonumber
     \\
     &\times& \sqrt{\pi}{\rm e}^{z\left(\tilde{t}\,\right)^2}{\rm erfc}\,\left[z\left(\tilde{t}\,\right)\right],
 \end{eqnarray}
 where $\tilde{\Gamma} \equiv \Gamma T_R$ and $g_0 \equiv 4|\gamma| E_0 T_R/\hbar$ is a dimensionless parameter. The pumping strength is the function that drives the squeezing processes, and directly affects the amount of squeezing in the ring. A large peak value in the pumping strength will generate substantial quadrature squeezing. In Fig. \ref{fig:e3time}, the pumping strength in the ring is plotted as a function of time for $a = \sigma = 0.99$ (critical coupling) and $g_0= .413$. Initially ($t=-\infty$), the pumping strength in the ring is zero. As the input pulse starts to coupling into the ring the pumping strength begins to build up. At $t=0$, the input pulse takes on its peak value at the coupling point in the channel. Some time later the pumping strength reaches its peak value. As can be seen, this time and the maximum value that  the pumping strength reaches depend on the duration of the input pulse $\tau$ in the channel. For a short input pulse duration of $\tilde{\tau} = 1$, the pumping strength very quickly builds up to its peak value. The longer the input pulse becomes, the more time it takes for this to occur. For very long input pulses, the peak pumping strength will scale as $1/\sqrt{\tau}$, but the dependence on the pump duration is more complicated for shorter pulses and as can be seen, the maximum pumping strength is in fact achieved for intermediate pulse durations. We denote the input pulse duration that maximizes the peak pumping strength by $\tau_g$. In the Appendix \ref{Doftaug}, we derive the following approximate but accurate expression for $\tau_g$ in the low-loss limit $(1-\sigma a)\ll 1$:
 \begin{equation}
    \label{taup_text}
    \tilde{\tau}_g\approx0.342\frac{\sqrt{8\ln2}}{1-\sigma a}.
 \end{equation}
Also in Appendix \ref{Doftaug}, we show that a pulse duration of $\tau_g$ given in Eq. \eqref{taup_text} causes the pumping strength to peak at the time
 \begin{equation}
     \label{tpeak}
      \tilde{t}_{peak} = \frac{1}{2(1-\sigma a)},
 \end{equation}
which is $1/\tilde{\Gamma}$, assuming that $(1-\sigma a)\ll 1$.
 
Before proceeding, we note that we could have used the field $E_4(t)$ rather than $E_3(t)$ and produced similar results. They are related by $E_3(t) = a E_4(t-T_R)$.  However, the field $E_3(t)$ is a more conservative representation of the field inside the ring, because it has been reduced by the attenuation loss of one additional round trip relative to $E_4(t)$.    

The initial conditions for equations \eqref{eq:sqamp} to \eqref{eq:thmnum} are evaluated at an early time, $t_i$ $(<0)$, when the incident pump pulse amplitude is negligible. The initial state of the system is the vacuum state, which means that $u(t_i) =0$ and $n_{\rm th}(t_i)=0$. We set the initial squeezing phase, $\phi(t_i)$, to be $\phi(t_i) = 0$, so that the time-dependent phase is given by $\phi(t) = -\omega_P (t-t_i)$. In numerical calculations, the absolute value of the initial time must be chosen such that  $|t_i| \gg \tau$.

The numerical solution of the coupled equations \eqref{eq:sqamp} to \eqref{eq:thmnum} enable us to determine the time-dependent level of quadrature squeezing in the ring. To this end, quadrature operators $\hat{X}$ and $\hat{Y}$  are defined as,
\begin{eqnarray}
\label{quadopx}
\hat{X} &=& \hat{b}^\dagger{\rm e}^{-i\theta(t)} + \hat{b}{\rm e}^{i\theta(t)},
\\
\label{quadopy}
\hat{Y} &=& i\left(\hat{b}^\dagger{\rm e}^{-i\theta(t)} - \hat{b}{\rm e}^{i\theta(t)}\right).
\end{eqnarray}
Here the quadrature phase $\theta(t)$ is defined as $\theta(t) \equiv \omega_S ( t- t_i)$. We include this phase so that the expectation value of the quadrature does not contain fast oscillations in time, because this choice cancels with the phase $\phi(t)$ of the squeezed state. The noise in the $\hat{X}$ and $\hat{Y}$ quadratures is defined as the square root of the variance, and written as $\Delta X$ and $\Delta Y$. Using Eq. \eqref{eq:solution} they can be shown to be given by, \cite{KimPropertiesStates}
\begin{eqnarray}
\Delta X(t) &=& \sqrt{2n_{\text{th}}(t) + 1}\;{\rm e}^{-u(t)} \label{eq:dx},
\\
\Delta Y(t) &=& \sqrt{2n_{\text{th}}(t) + 1}\;{\rm e}^{u(t)}. \label{eq:dy}
\end{eqnarray}
 Multiplying Eqs. \eqref{eq:dx} and \eqref{eq:dy} together gives, 
 \begin{equation}
     \label{nthdxdy}
     \Delta X (t) \Delta Y (t) = 2n_{\text{th}}(t)+ 1. 
 \end{equation}
If $n_{\rm th} = 0$, then $\Delta X \Delta Y = 1$ and a squeezed vacuum state is recovered, with $\Delta X = \exp(-u)$ and $\Delta Y = \exp(u)$. With our choice of quadrature operators, the noise in either quadrature for a vacuum state ($u=0$) is simply $\Delta X  = 1$ and $\Delta Y = 1$. Therefore, squeezing below the vacuum noise in the $\hat{X}$ quadrature occurs when $\Delta X < 1$ in Eq. \eqref{eq:dx}. The expectation value of the photon number for the squeezed thermal state can be shown to be given by \cite{KimPropertiesStates}
\begin{equation}
    \label{eq:totalphoton}
  \left<\hat{n}\right>\equiv\left<\hat{b}^\dagger \hat{b}\right> =   n_{\rm th}(t)\cosh\left(2u(t)\right) + \sinh^2\left(u(t)\right).
\end{equation}
When $n_{\rm th} = 0$, the expectation value of the photon number is $\sinh^2(u)$, which is the result obtained for a squeezed vacuum state.

\section{Results and Discussion}
\label{sec:results}
In this section, we present our numerical solutions to  the set of equations \eqref{eq:sqamp} to \eqref{eq:thmnum}. We solve them using a fourth-order Runge-Kutta method; the total run time for a given configuration is on the order of a few seconds on a standard PC. We also derive an approximate analytic expression for the minimum quadrature noise in terms of the peak pumping strength, and an expression for the optimum choice of $\sigma$ (or alternatively, $\kappa$) that produces the global minimum in the quadrature noise. In addition, we numerically determine the pulse duration that produces the minimum quadrature noise for a given $\kappa$ and show that it is close to $\tau_g$, as given in Eq. \eqref{taup_text}.  We discuss the effects of scattering loss $a$ on the quadrature noise, and the optimum coupling coefficient and pulse duration. Finally, we study the sensitivity of the minimum quadrature noise to a phase offset due to imperfect homodyne detection.

In the remainder of this paper, we use the following values for our pump and ring parameters. We take the ring material to be AlGaAs with  $\chi^{(2)}_{\rm eff} = 100\, {\rm pm}/{\rm V}$ \cite{Yang2007GeneratingResonators}, $n_{\rm eff} = 2.85$, and $\omega_P = 2\pi\times135.73\,{\rm THz}$ ($\lambda_P = 775\, {\rm nm}$). The amplitude of the input pulse $E_0$ can be written in terms of the total pump pulse energy $U$ as,
\begin{equation}
    \label{E0}
    E_0 = \left(\frac{4\ln2}{\pi}\right)^{1/4}\sqrt{\frac{2U}{\mathcal{A}c\, n_{\rm eff} \epsilon_0  T_R}},
\end{equation}
 where $\mathcal{A} = 0.71\, {\mu {\rm m}}^2$ is the cross-sectional area of the ring waveguide, $\epsilon_0$ is the permittivity of free space, and $c$ is the speed of light. The energy of the incident pulse is chosen to be $U = 0.188\,{\rm pJ}$ (independent of the pulse duration). This value of $U$ produces a substantial amount of squeezing, but generally does not lead to significant pump depletion, even for low-loss cavities.

The radius of the ring required to give a resonance at the pump frequency is $R = m_P c/ (\omega_P n_{\rm eff})$, where we have used Eq. \eqref{eq:thephase} with $\omega = \omega_P$ and $\Theta = 2\pi m_P$. We choose the pump mode number to be $m_P=200$, which makes the ring radius approximately equal to $R \approx 25\,{\rm \mu m} $. The ring round trip time is given by $T_R = 2\pi R n_{\rm eff}/c$, and in this case is $T_R \approx 1.47 \, {\rm ps} $. We present our results in terms of the dimensionless parameters; $\tilde{t}\equiv t/T_R$ and $\tilde{\tau}\equiv \tau/T_R$. Once this is done, the only place where $R$ enters our model is in the amplitude of the pumping strength in Eq. \eqref{pumpapprox}. Thus in order to make our results independent of $R$ we require  $E_0 T_R$ be constant. We collect all the dimensional parameters above into the single dimensionless constant, $g_0$, which was introduced in Eq. \eqref{pumpapprox}. For the above choice of parameters, $g_0 = 0.413$.

\subsection{Dynamics of the squeezing process}

\begin{figure}[htbp]
        \centering
        \begin{subfigure}[b]{.475\textwidth}
            \centering
            \includegraphics[width=\textwidth]{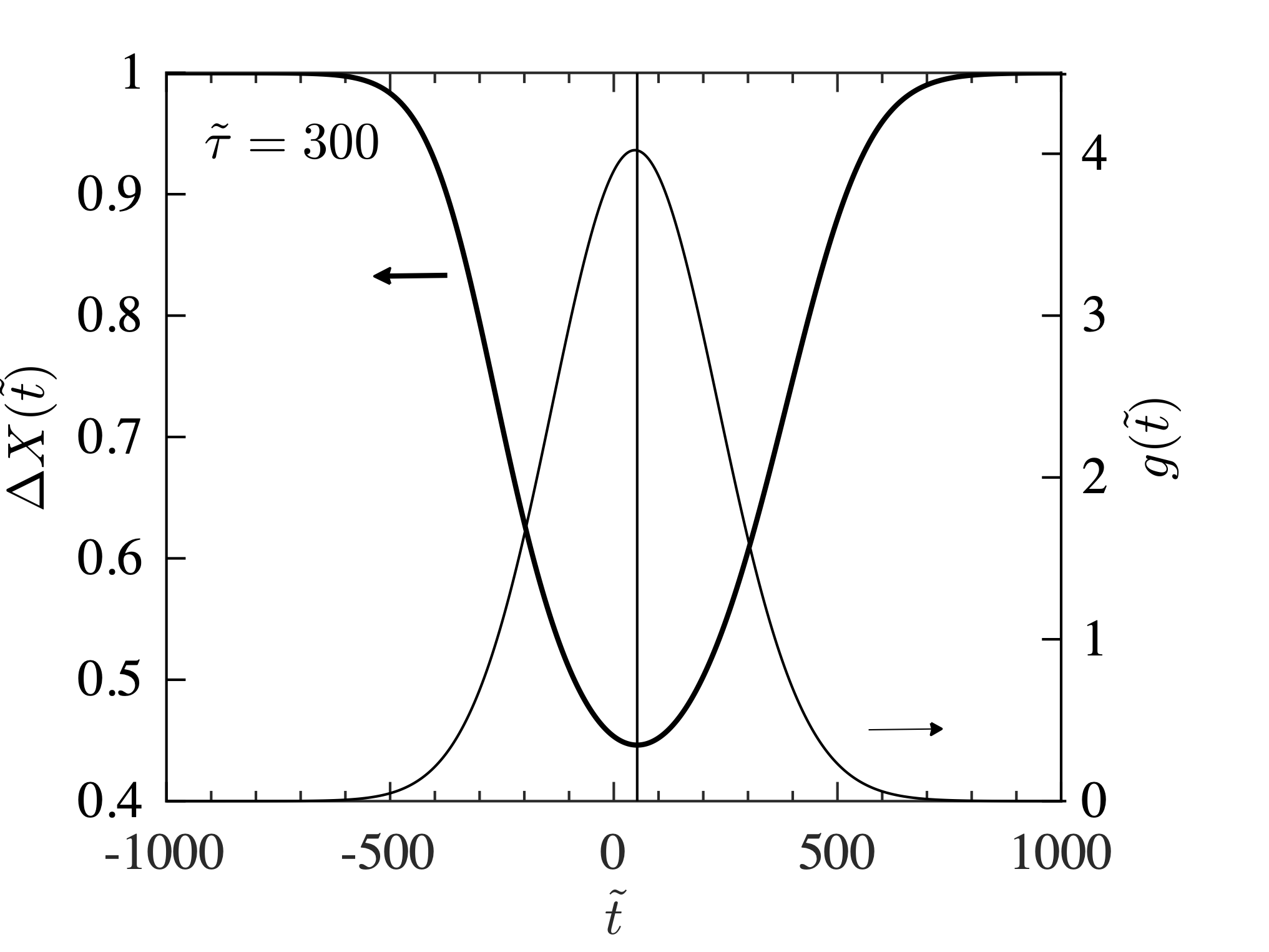}
            \caption[]%
            
            \label{fig:dxandg300}
        \end{subfigure}
        \vfill
        \begin{subfigure}[b]{.475\textwidth}  
            \centering 
            \includegraphics[width=\textwidth]{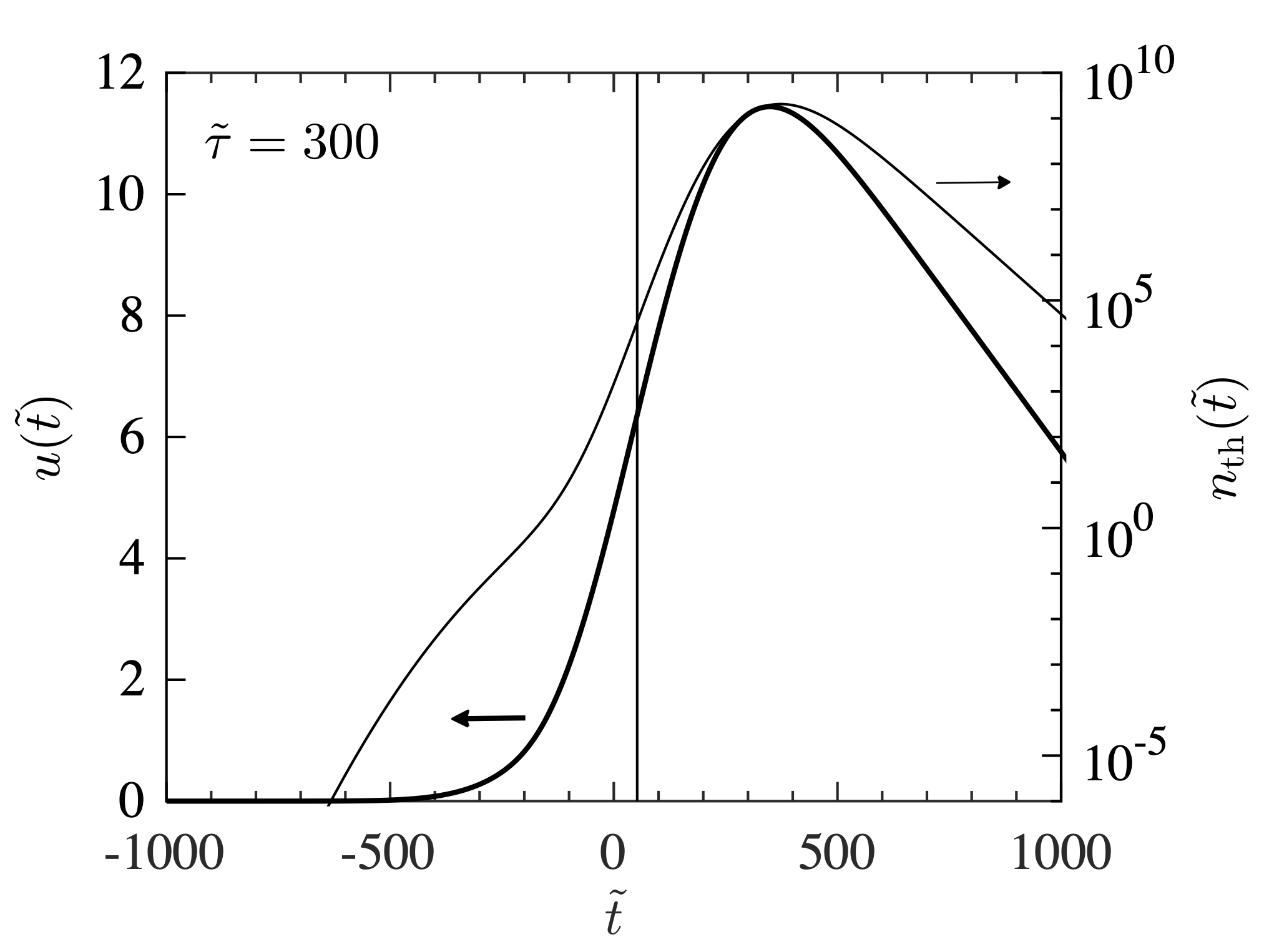}
            \caption[]%
            
            \label{fig:sqandnth300}
        \end{subfigure}
      \caption{\small (a) The quadrature squeezing $\Delta X$ (thick line) and pumping strength $g$ (thin line) as a function of time, and (b) the squeezing amplitude $u$ (thick line) and thermal photon number $n_{\rm th}$ (thin line) as a function of time for an input pulse duration of $\tilde{\tau}=300$ and coupling constant $\sigma = a =0.99$. The time at which $\Delta X$ is minimum is $\tilde{t}_{\rm min}\approx 40$ is indicated by the vertical line.}
\label{fig:tau300dynamics}
\end{figure}

We begin by examining the time-dependent quadrature noise $\Delta X$ in the ring in Fig. \ref{fig:dxandg300} for $\sigma = a = 0.99$ (critical coupling), for an input pulse duration of $\tilde{\tau}=300$. Initially the pumping strength is zero and the quadrature noise is equal to the vacuum noise $\Delta X = 1$. As the pumping strength builds up, the quadrature noise gets squeezed below the vacuum noise $\Delta X < 1$. We find that the quadrature noise is a minimum at approximately (but not exactly) the time at which the pumping strength is at its peak, that is, at $\tilde{t}_{\rm min}\approx 40$ (indicated by the vertical line). Finally, when the pump pulse couples out of the ring, the quadrature noise returns to the vacuum noise. The time-dependent squeezing amplitude $u$ and thermal photon number $n_{\rm th}$ are shown in Fig. \ref{fig:sqandnth300} for the same parameters.  As the squeezing amplitude increases, the quadrature noise is  squeezed by the factor $\exp(-u)$. However, the trade off is that the thermal photon number also increases, which results in an increase in the quadrature noise by the factor $\sqrt{2n_{\rm th} +1}$. Thus the minimum quadrature noise does not happen when the squeezing amplitude is maximum, but instead at an earlier time closer to when the pumping strength is maximum and the thermal photon number is much less than its peak value. 

\begin{figure}[htbp]
        \centering
        \begin{subfigure}[b]{.475\textwidth}
            \centering
            \includegraphics[width=\textwidth]{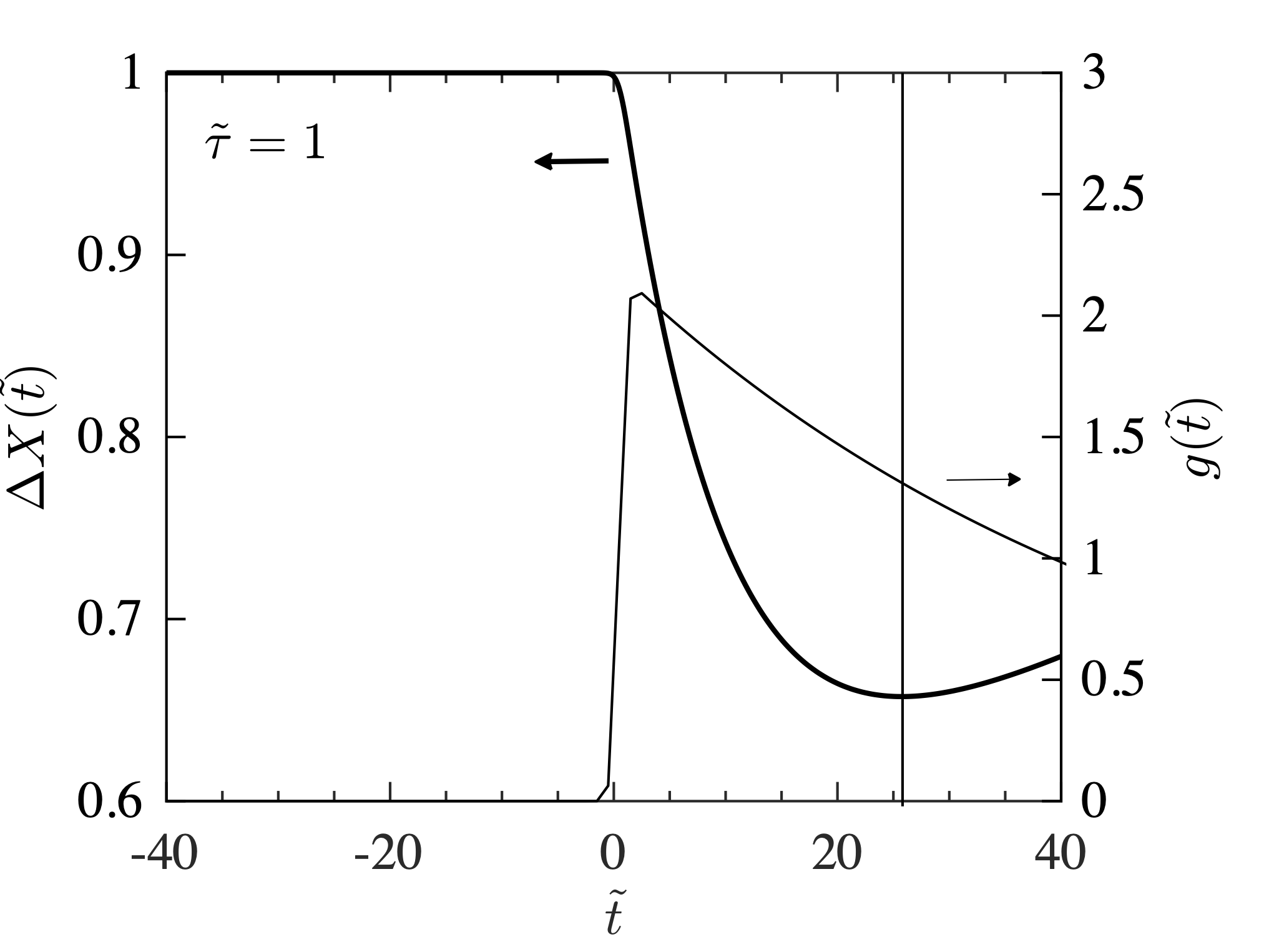}
            \caption[]%
            
            \label{fig:dxandg1}
        \end{subfigure}
        \vfill
        \begin{subfigure}[b]{.475\textwidth}  
            \centering 
            \includegraphics[width=\textwidth]{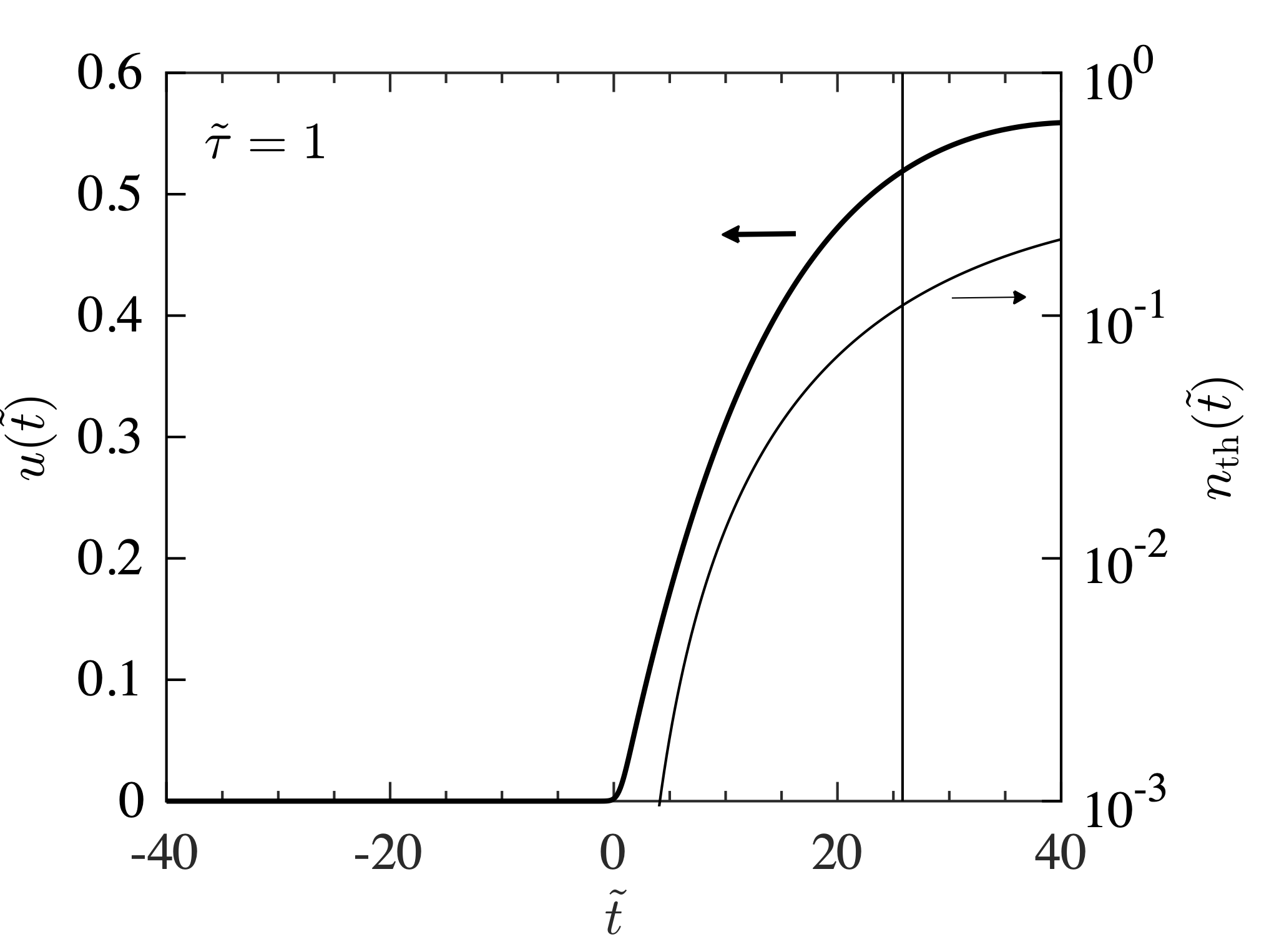}
            \caption[]%
            
            \label{fig:sqandnth1}
        \end{subfigure}
      \caption{\small The same plots as in Fig. \ref{fig:tau300dynamics} but for an input pulse duration of $\tilde{\tau}=1$. Note that now the time at which $\Delta X$ is minimum is $\tilde{t}_{\rm min}\approx 26$.}
\label{fig:tau1dynamics}
\end{figure}

In Fig. \ref{fig:dxandg1}, we examine a similar setup as above, except our input pump pulse has a much shorter duration of $\tilde{\tau} = 1$. Here, the pumping strength quickly reaches its peak value and does not spend much time building up in the ring. The quadrature noise is not as squeezed as it was with the long pulse. Additionally, with the short pulse, the minimum quadrature noise does not occur at the same time as when the pumping strength is at its peak. In this case the peak pumping strength occurs at approximately $\tilde{t}\approx 2$ and the minimum quadrature noise occurs at approximately $\tilde{t}_{\rm min}\approx 26$.  The time-dependent squeezing amplitude and thermal photon number are shown in Fig. \ref{fig:sqandnth1} for the same short pulse. The thermal photon number is significantly smaller now, so the factor $\sqrt{2n_{\rm th} +1}$ is less detrimental to the squeezing. As a result we find that the minimum quadrature noise now occurs closer to the time when the squeezing amplitude is at its peak value.

\begin{figure}[htbp]
        \centering
        \begin{subfigure}[]{.475\textwidth}
            \centering
            \includegraphics[width=\textwidth]{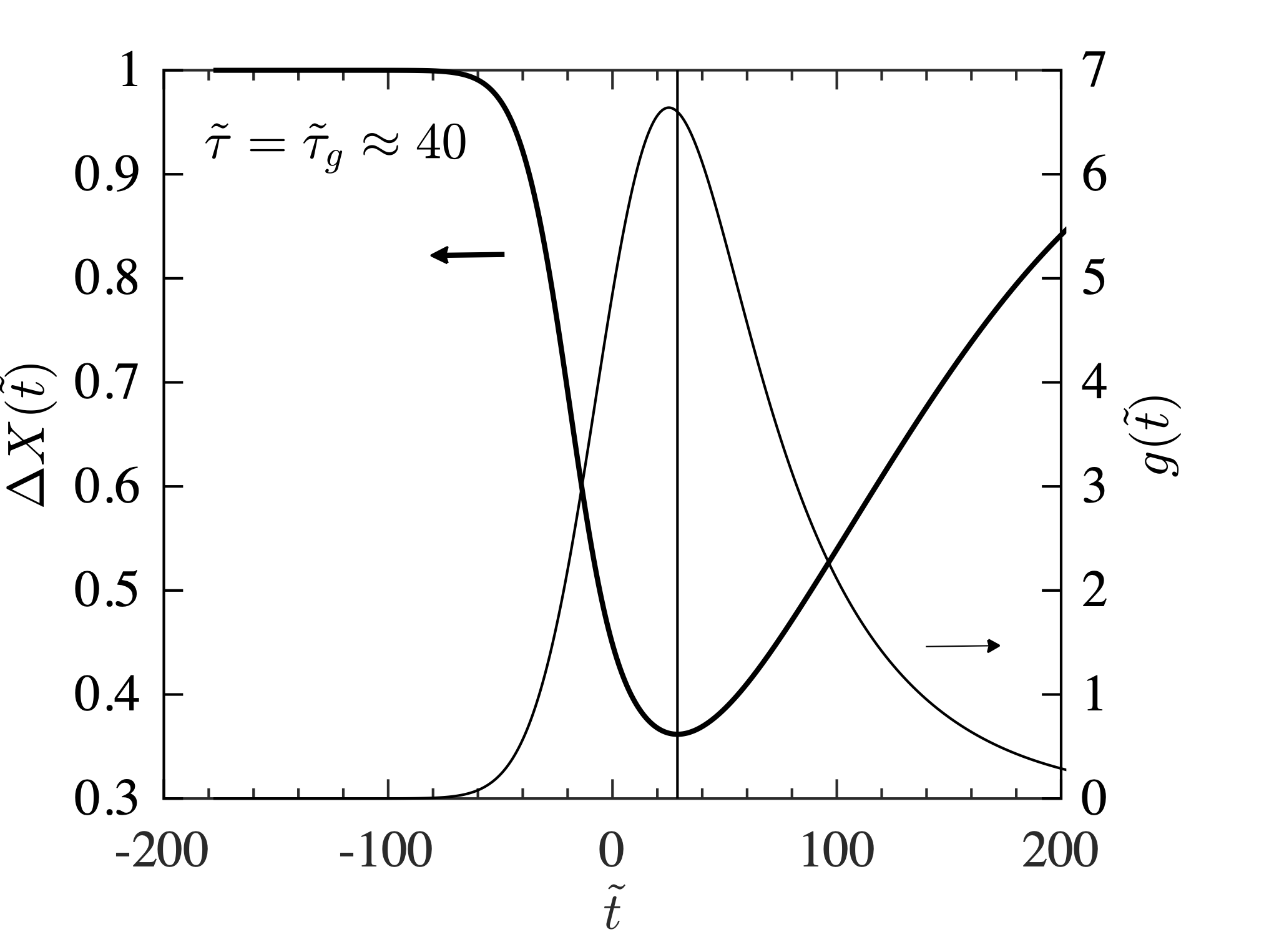}
            \caption[]%
            
            \label{fig:dxandg40}
        \end{subfigure}
        \begin{subfigure}[]{.475\textwidth}  
            \centering 
            \includegraphics[width=\textwidth]{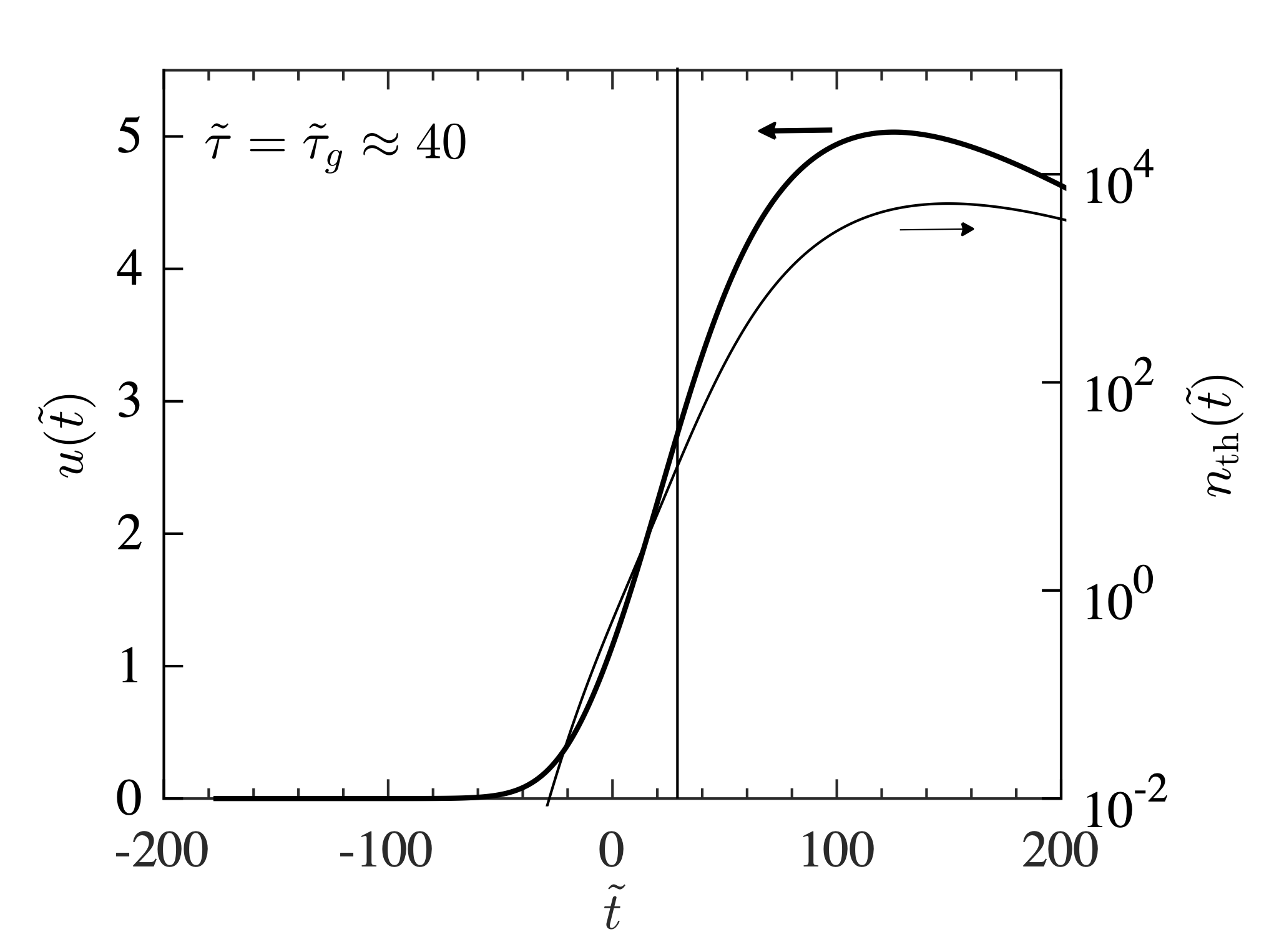}
            \caption[]%
           
            \label{fig:sqandnth40}
        \end{subfigure}
      \caption{\small The same plots as in Fig. \ref{fig:tau300dynamics} but for an input pulse duration of $\tilde{\tau}=\tilde{\tau}_g\approx 40$. Note that now the time at which $\Delta X$ is minimum is $\tilde{t}_{\rm min}\approx 29$.}
\label{fig:tau40dynamics}
\end{figure}

Having examined the two extreme cases of a long pulse and short pulse, we now consider the most interesting case for quadrature squeezing. We pump the ring with an input pulse duration $\tau_g$ (given by Eq. \eqref{taup_text}) that gives the greatest peak value of the pumping strength. For $\sigma = a = 0.99$, $\tilde{\tau}_g\approx 40$. In Fig. \ref{fig:dxandg40} the time-dependent quadrature noise is shown for this pulse. When compared to the short and long pulse, we find that this duration produces the greatest quadrature squeezing. The minimum quadrature noise occurs at roughly the same time as the peak value of the pumping strength; using Eq. \eqref{tpeak} the peak pumping strength occurs at $\tilde{t}_{peak}\approx 25 $, while the quadrature noise is a minimum, at $\tilde{t}_{\rm min}\approx 29$. The time-dependent squeezing amplitude and thermal photon number for this pulse duration are shown in Fig. \ref{fig:sqandnth40}. The peak squeezing amplitude is reduced by a factor of approximately 2 compared to the long pulse. However, the depletion of the squeezing amplitude is counteracted by the thermal photon number being reduced by a factor of roughly $10^5$. This shows that the thermal noise is much more sensitive to the duration of the input pulse than the squeezing amplitude is, and therefore it is better to err on the side of using a relatively shorter pulse in a lossy ring resonator. 

\subsubsection{Minimum in the quadrature noise}

We have demonstrated how the minimum in $\Delta X$ depends on the pulse duration $\tau$. Here we derive an analytic expression for the minimum quadrature noise $\Delta X_{\rm min}$. Setting the derivative of $\Delta X (t)$ in Eq. \eqref{eq:dx} equal to zero at the time $t_{\rm min}$ and simplifying gives,
 \begin{equation}
 \label{dxdiff02}
     \frac{d n_{\rm th}\left(t\right)}{dt}\bigg|_{t=t_{\rm min}}-
     \left(2 n_{\rm th}\left(t_{\rm min}\right)+1\right)\frac{d u\left(t\right)}{d t}\bigg|_{t=t_{\rm min}} = 0.
 \end{equation}
  Replacing the derivatives in Eq. \eqref{dxdiff02} with   Eq. \eqref{eq:sqamp} and Eq. \eqref{eq:thmnum} and using Eq. \eqref{eq:dx} to simplify gives,
 \begin{eqnarray}
     \label{dxdiff04}
     \Delta X_{\rm min}\left(\tau\right) = \frac{1}{\sqrt{1+g(t_{\rm min},\tau)}},
 \end{eqnarray}
 where $g(t_{\rm min},\tau)$ is the pumping strength evaluated at the time when the quadrature noise is at its minimum. In general, we evaluate $g(t_{\rm min},\tau)$ numerically in order to calculate the minimum quadrature noise for a given $\sigma$, $a$, and $\tau$. If the input pulse duration is close to or larger than $\tau_g$, then the value of the pumping strength at the time when the quadrature noise is minimum is roughly the same as the peak value of the pumping strength (see Figs. \ref{fig:dxandg300} and \ref{fig:dxandg40}). Thus, we can neglect the difference between $g(t_{\rm min})$ and the peak value of the pumping strength. That is, if the pulse duration is considerably longer than $T_R$, then the pumping strength does not vary appreciably over a time scale of a few round-trips of the ring. This approximation improves the longer the pulse.  Conversely, this approximation is not valid for the setup in Fig. \ref{fig:dxandg1} for the short pulse, as we discussed earlier. Let $g_{\rm max}(\tau)$ denote the peak pumping strength as a function of $\tau$. Then, since for pulses durations $\tau \gg T_R$ $g_{\rm max}(\tau)\approx g(t_{\rm min},\tau)$, we obtain the following approximate expression for the minimum quadrature noise:
 \begin{equation}
     \label{dxapprox}
     \Delta X_{\rm min}(\tau) \approx \frac{1}{\sqrt{1+g_{\rm max}(\tau)}}, \,\,\,\, (\tau\gtrsim \tau_g).
 \end{equation}
 Therefore the minimum quadrature noise is expressed in terms of the peak pumping strength, for which we have an expression in Eq. \eqref{pumpapprox}. The advantage of Eq. \eqref{dxapprox} is that it gives the minimum quadrature noise as a function of $\tau$ and $\sigma$; without having to solve the coupled differential equations numerically. Additionally, letting $\tau = \tau_g$ in Eq. \eqref{dxapprox}, and using Eq. \eqref{taup_text} and Eq. \eqref{tpeak}, gives the following result,
 \begin{equation}
     \label{dxapproxtaup}
     \Delta X_{\rm min}(\tau_{g}) \approx \left[1+ 0.653 \, \frac{g_0a }{\tilde{\Gamma}}\sqrt{\frac{1-\sigma^2}{1-\sigma a}}\right]^{-\frac{1}{2}}.
 \end{equation}
This is the minimum quadrature noise in the ring for the pulse duration of $\tau_g$, as a function of $\sigma$ and $a$. For a given $\sigma$ and $a$, we will show in the next section that this expression approximately gives the best quadrature squeezing. We will assess the accuracy of the expression given in Eqs. \eqref{dxapprox} and \eqref{dxapproxtaup} below.

\subsection{Dependence of the minimum quadrature noise on pulse duration and coupling}

The minimum quadrature noise depends on the pulse duration $\tau$, coupling $\sigma$, and scattering loss $a$. Thus far, the numerical results that we have presented have been only for the case of very low scattering loss at critical coupling ($\sigma =a = 0.99$), and for only three pulses. We have shown that, compared to a short and long pulse, $\tau_g$ generates the best quadrature squeezing for a given $\sigma$ and $a$. In this section, we present numerical results for the maximum quadrature squeezing as a function of the coupling constant and pump duration for different scattering loss in the ring.  We will show that the choice of critical coupling, although an obvious starting point, is not the optimal choice in order to achieve the global minimum in the quadrature noise for a given $a$. In fact, we find the global minimum in the quadrature noise is in the undercoupled ($\sigma > a$) regime and derive an approximation analytic expression for the optimal coupling. 

Our analysis is done by computing the minimum quadrature noise $\Delta X_{\rm min}(\tau,\kappa)$ as a function of pulse duration and coupling for different attenuation constants $a$. Then we numerically determine the optimal choices for the pulse duration and coupling, and finally compare them to approximate analytic expressions that we derive. 

\begin{figure*}[htbp]
        \centering
        \begin{subfigure}[b]{0.49\textwidth}
            \centering
            \includegraphics[width=\textwidth]{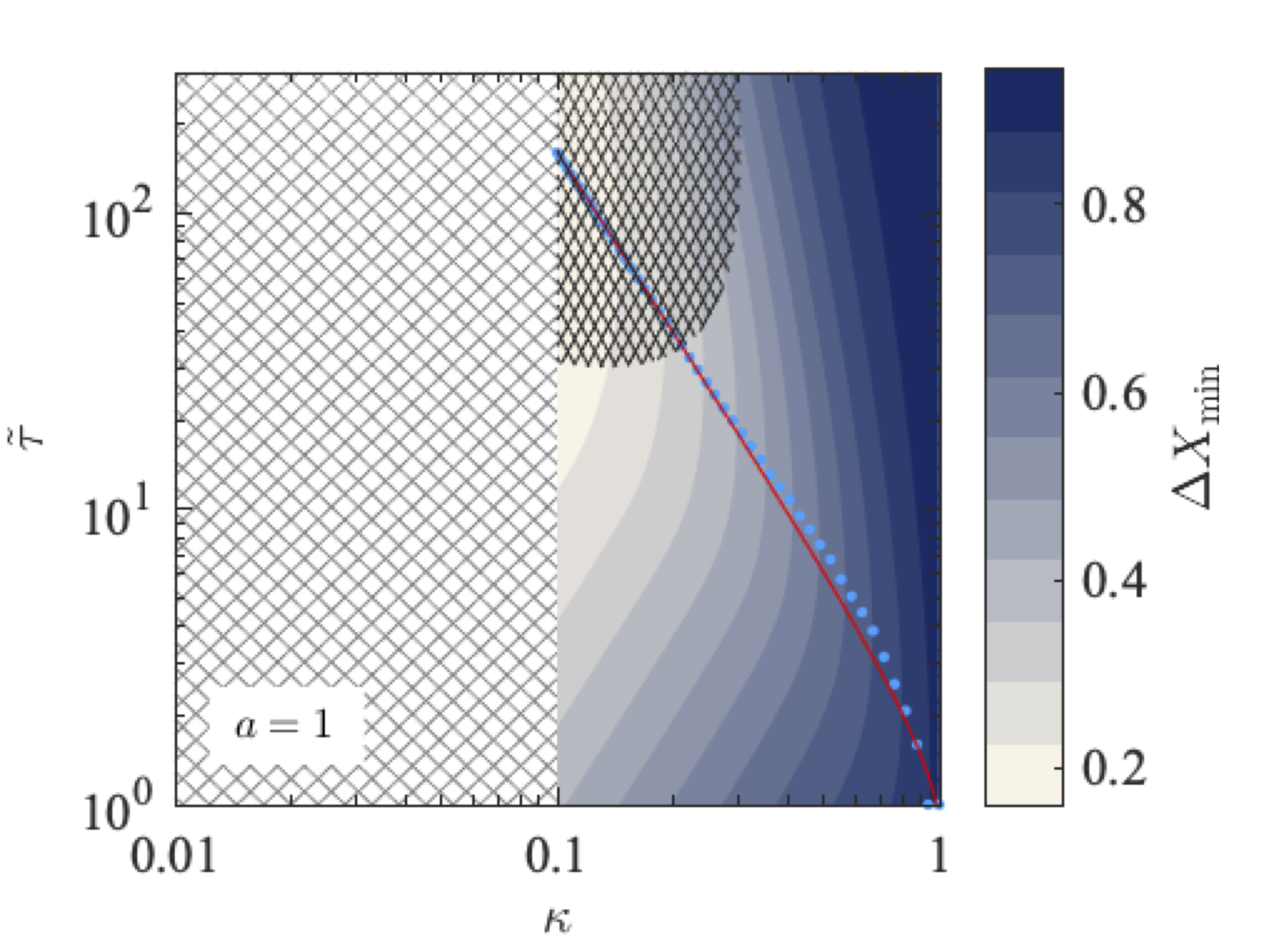}
            \caption[]%
            
            \label{fig:a1}
        \end{subfigure}
        \hfill
        \begin{subfigure}[b]{0.49\textwidth}  
            \centering 
            \includegraphics[width=\textwidth]{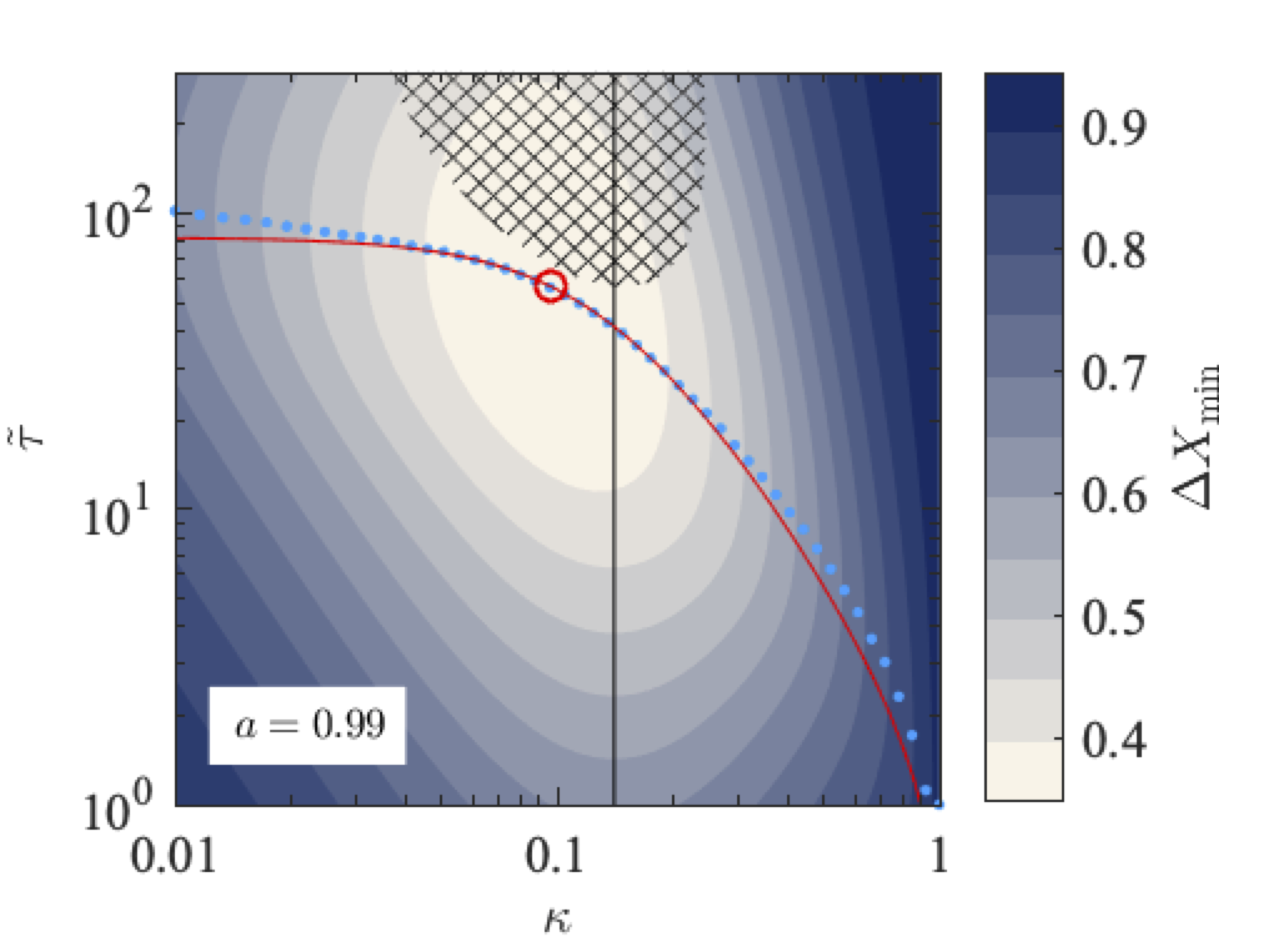}
            \caption[]%
            
            \label{fig:a99}
        \end{subfigure}
        \vskip\baselineskip
        \centering
        \begin{subfigure}[b]{0.485\textwidth}   
            \centering 
            \includegraphics[width=\textwidth]{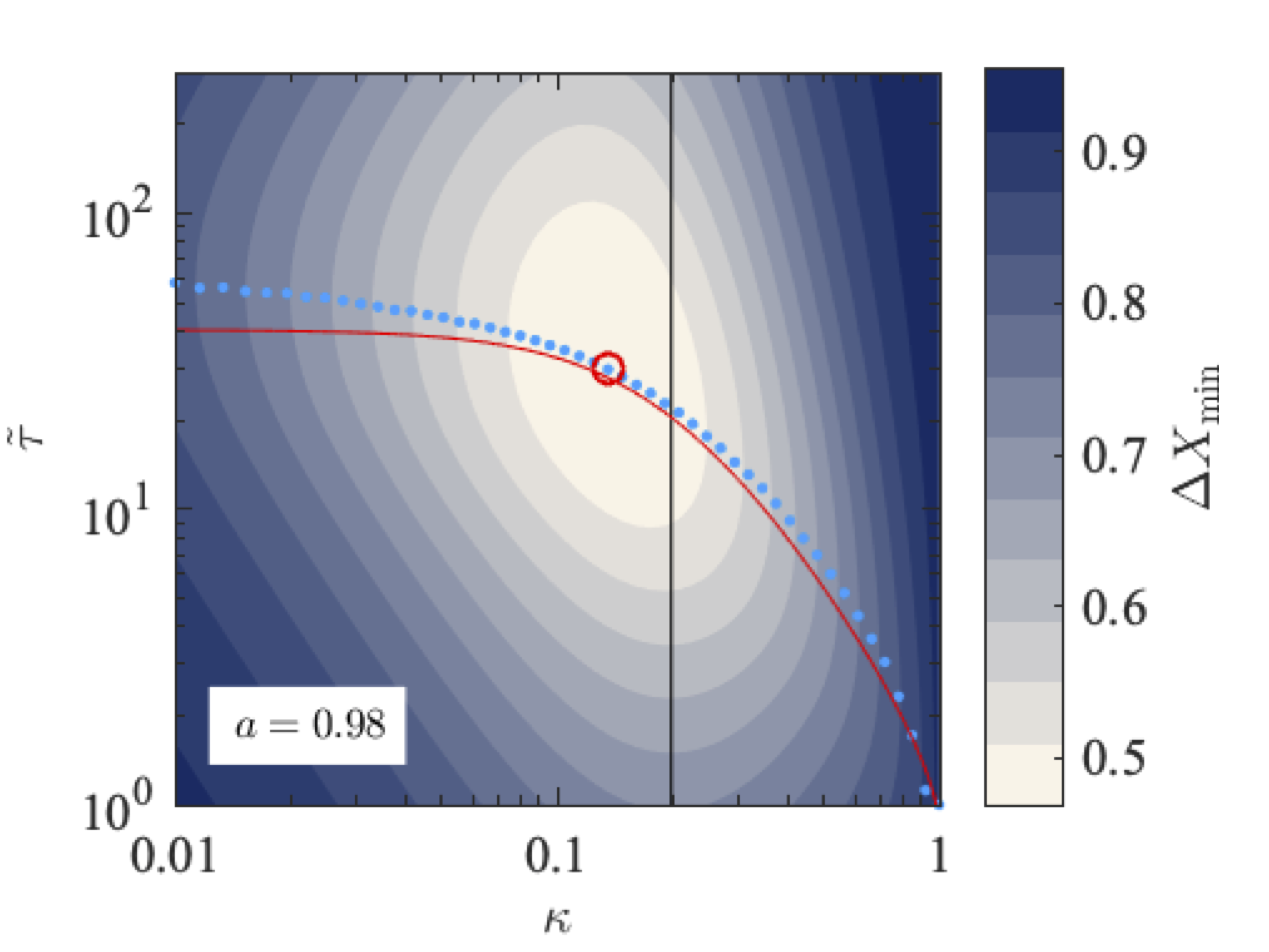}
            \caption[]%
  
            \label{fig:a98}
        \end{subfigure}
        \quad
        \begin{subfigure}[b]{0.485\textwidth}   
            \centering 
            \includegraphics[width=\textwidth]{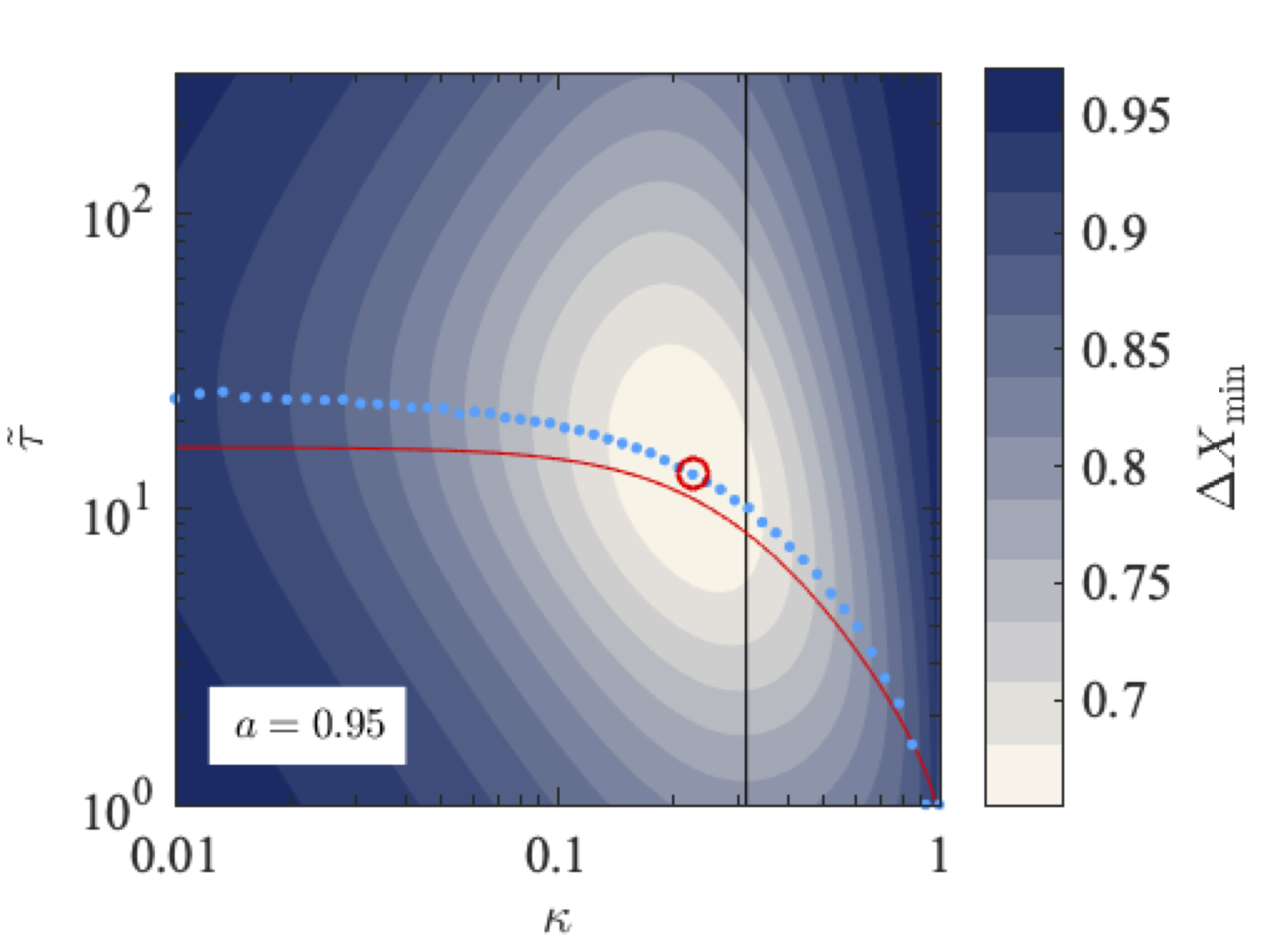}
            \caption[]%
            
            \label{fig:a95}
        \end{subfigure}
        \caption[]
        {\small The minimum quadrature noise $\Delta X_{\rm min}$ as a function of the input pulse duration $\tilde{\tau}$ and cross-coupling constant $\kappa$ for an attenuation constant of (a) $a=1$, (b) $a=0.99$, (c) $a=0.98$, and (d) $a=0.95$. The blue dots indicate the computed pulse duration  needed to minimize $\Delta X_{\rm min}$ for a given $\kappa$. The solid red line is the pulse duration $\tilde{\tau}_g(\kappa)$ as a function of $\kappa$ given by Eq. \eqref{taup_text}. The red circles in (b)-(d) mark the point at which the quadrature noise is at a global minimum for the given value of $a$. The vertical black line in (b)-(d) indicates critical coupling ($\sigma =a$, \textit{i.e.}, $\kappa = \sqrt{1-a^2}$). The light hatched area in (a) marks the parameter space where our simulation does not converge. The dark hatched areas in (a) and (b) indicate regions where the number of generated photons is in excess of $1\%$ of the of photons in the incident pump.} 
        \label{fig:minDX3d}
\end{figure*}

 In Fig. \ref{fig:minDX3d} we plot the minimum quadrature as a function of the coupling coefficient and pulse duration for four different loss parameters $a$. First, in Fig. \ref{fig:a1}, we consider the case where there is no scattering loss ($a=1$). In this case, the minimum quadrature noise decreases as the cross-coupling constant $\kappa$ goes to zero (or $\sigma$ goes to one). Consequently, we find no optimum value of $\kappa$ that gives a global minimum in the quadrature noise. This is expected, because as $\kappa$ goes to zero, the buildup factor continues to increase without bound. In the figure, there are two hatched areas. The darker hatching (around $\kappa = 0.1$) is where the expectation value of the number of generated photons is at least $1\%$ of the average number of pump photons ($\left<n_{\rm pump}\right>\sim 2\times 10^6$); thus, our undepleted pump approximation is becoming less accurate. The second, lighter hatching (where $\kappa < 0.1$) indicates when our simulations break down, because the decay rate $\tilde{\Gamma}$ goes to zero as $\kappa$ goes to zero. The blue dotted line in the figure indicates the computed pulse duration that gives the best quadrature squeezing for a given $\kappa$. The red curve is the input pulse duration $\tau_g(\kappa)$ given by Eq. \eqref{taup_text}. The fact that $\tau_g$ fits agrees well with the computed optimal pulse duration means that the minimum in the quadrature noise is approximately where the peak pumping strength is the greatest. For short pulses, or pulses larger than $\tau_g$, the peak pumping strength is too small in the ring and we see that the squeezing gets worse.

We now consider how introducing scattering loss into the ring affects the squeezing. When there is loss, the buildup factor has a peak value at critical coupling $\kappa = \sqrt{1-a^2}$ (or $\sigma = a$). In Fig. \ref{fig:a99} the scattering loss is $a=0.99$. Consequently, there is substantial squeezing at the peak in the buildup factor at critical coupling (indicated by the vertical line), and the squeezing gets worse away from the peak, as $\kappa$ goes to zero (undercoupling) or one (overcoupling). We observe excessive photon generation, at least as much as $1\%$ of the average number of photons in the pump pulse (hatched area), for pulses longer than $\tau_g$ near critical coupling. The optimum squeezing point (indicated by the red circle) is at a $\kappa$ that is lower than critical coupling in the undercoupled regime, where the buildup of pump intensity is less, but the cavity decay rate is smaller. This shows that in order to achieve the largest squeezing it is preferable to have a lower cavity decay rate than that obtained at critical coupling. 

In Figs. \ref{fig:a98} and \ref{fig:a95} we increase the attenuation loss in the ring to $a=0.98$ and $a=0.95$, respectively. As the scattering loss in the ring is increased, critical coupling shifts to higher $\kappa$ and so does the optimum point (indicated by a red circle); however, it still remains in the undercoupled regime. In addition, the optimum point shifts to shorter pulses, which is expected, because the longer the pulse is, the more thermal photons are generated. Our approximate expression $\tau_g (\kappa)$ is still in quite good agreement with the numerical results, but is not as accurate as when the loss was very low ($a=0.99$). This is because it is an approximate expression that is valid only when $(1-\sigma a) \ll 1$ (see Appendix \ref{Doftaug}). Interestingly, it still fits quite well at the optimum coupling point, with a difference of less than $2.3 T_R$ or a relative error of $18\%$ when $a=0.95$. Using the approximate value for the pulse duration in this case only leads to a  $1\%$ increase in the quadrature noise relative to the optimal value.

\begin{figure}[h!]
     \centering
     \includegraphics[scale=0.95]{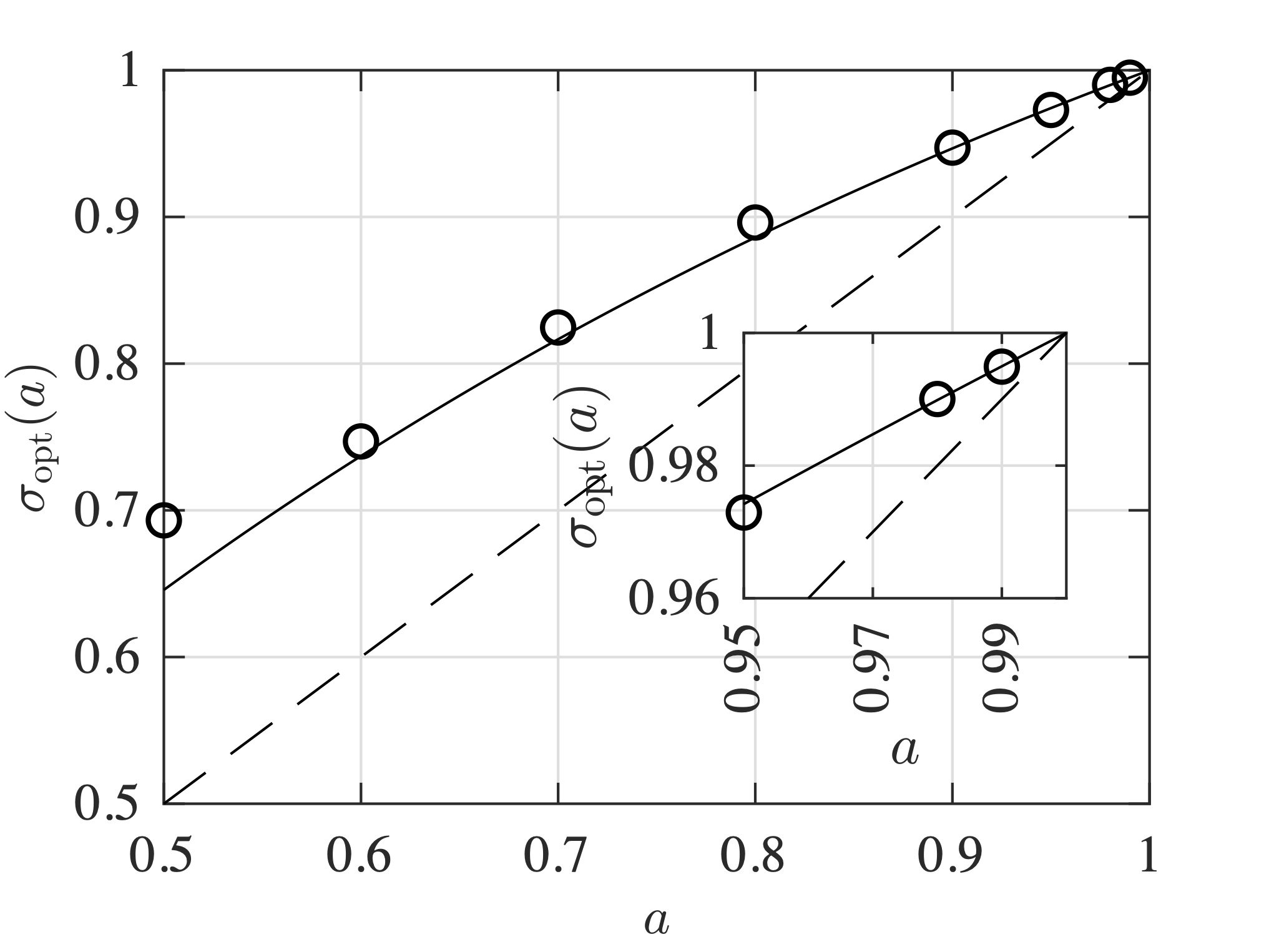}
     \caption{\small The computed optimum self-coupling constant (circles) and the approximate optimum coupling constant given by Eq. \eqref{eq:sigmaopt} (solid line), as a function of attenuation loss $a$. The dashed line indicates critical coupling.}
     \label{fig:optkappa}
 \end{figure}

An approximate expression for the optimum coupling value $\sigma_{\rm opt}$ (or $\kappa_{\rm opt}$) is given by minimizing $\Delta X_{\rm min} (\tau_g)$ in Eq. \eqref{dxapproxtaup} with respect to $\sigma$ for a fixed $a$. Doing this we obtain,
\begin{equation}
    \label{eq:sigmaopt}
    \sigma_{\rm opt}(a) \approx \frac{-1+\sqrt{3 a^2 +1}}{a}.
\end{equation}
This is a good approximation as long as $(1-\sigma a)\ll 1$ and $\tau\gtrsim \tau_g$. In Fig. \ref{fig:optkappa} we compare the $\sigma_{\rm opt}$ given by Eq. \eqref{eq:sigmaopt} (curve) to the numerically-computed value (circles). We find the analytic result fits well for $a \ge 0.9$. Note that as the scattering loss increases, the difference between critical coupling (dashed line) and $\sigma_{\rm opt}$ increases. Thus, for lossy systems the optimum coupling value $\sigma_{\rm opt}$ shifts closer to one (undercoupling) as compared to critical coupling. This compensates for the decrease in $a$ and makes the decay rate smaller. We note that the difference between the quadrature noise at critical coupling and optimum coupling is generally small; for $a=0.95$, the quadrature noise is reduced by only $\sim 0.3$ dB, and for $a=0.99$ by only $\sim 0.2$ dB  (see Figs. \ref{fig:a95} and \ref{fig:a99}). However, it is useful to know that one should err on the side of undercoupling if possible. 
\subsection{Comparing the analytic expression for the minimum quadrature noise to the numerical results }

\begin{figure}[htbp]
        \centering
        \begin{subfigure}[htbp]{.49\textwidth}
            \centering
            \includegraphics[width=\textwidth]{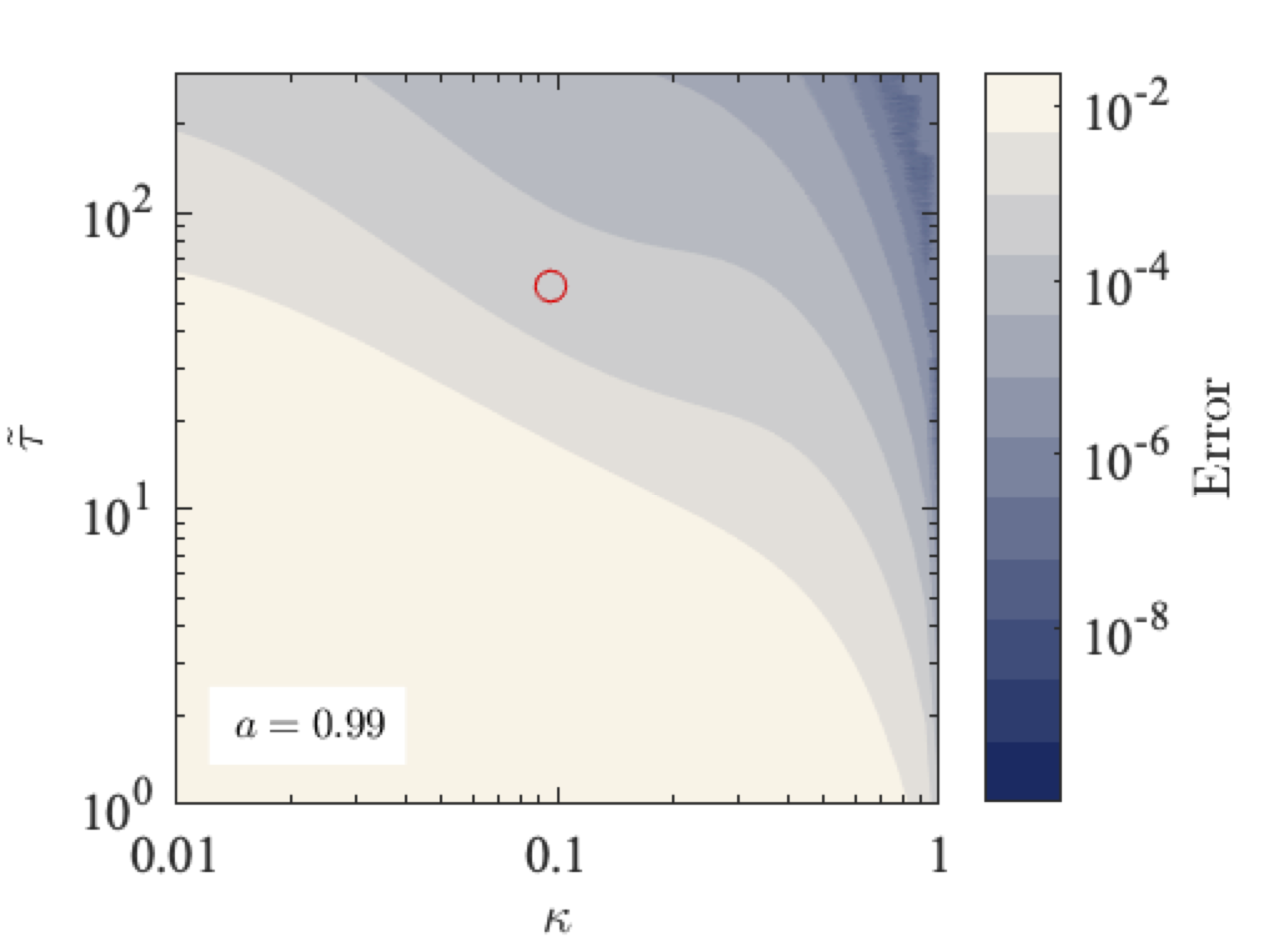}
            \caption[]%
            
            \label{fig:diff99}
        \end{subfigure}
        ~
        \begin{subfigure}[htbp]{.49\textwidth}  
            \centering 
            \includegraphics[width=\textwidth]{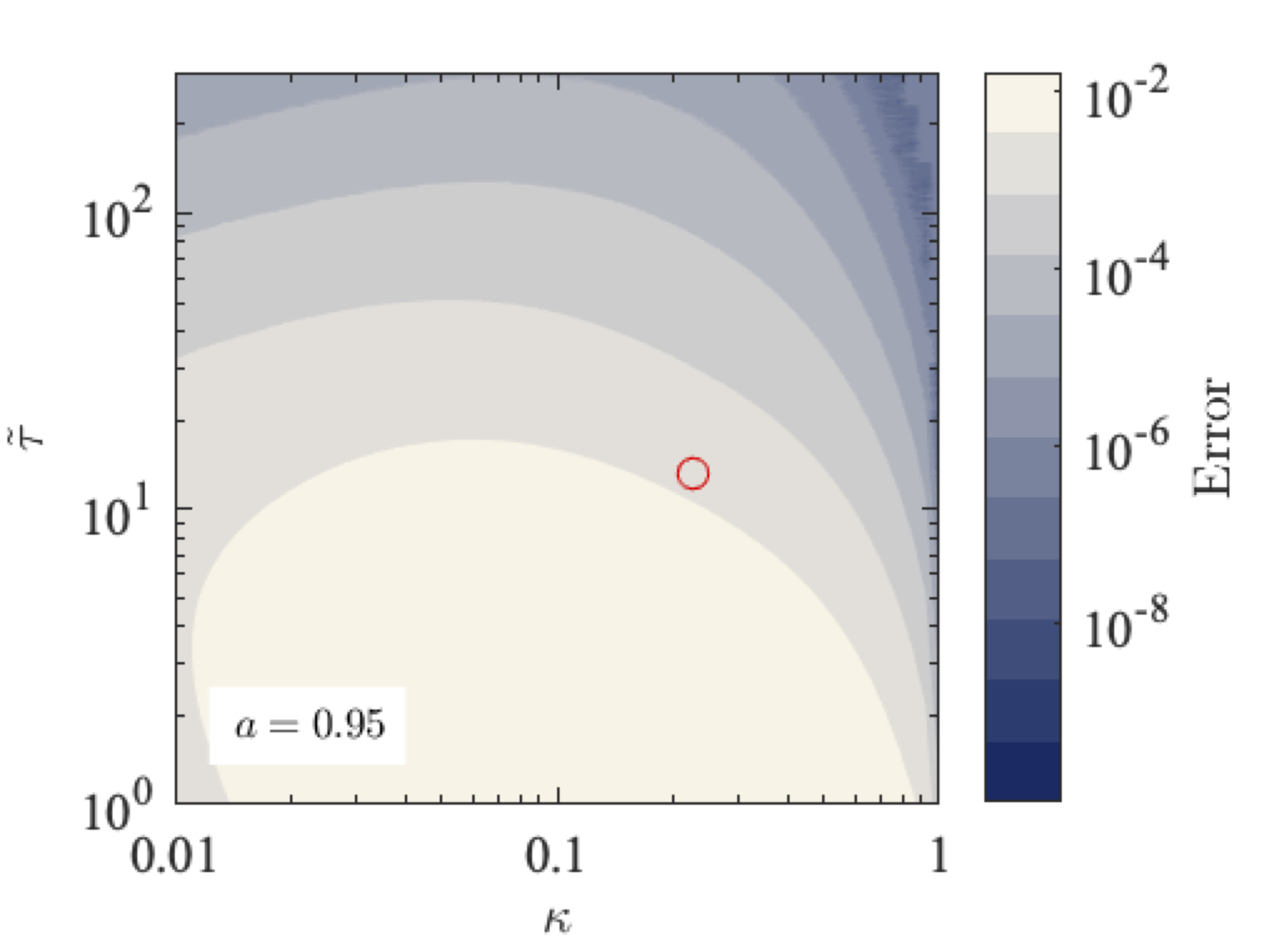}
            \caption[]%
            
            \label{fig:diff95}
        \end{subfigure}
        
        \caption{\small The absolute value of the relative error (see Eq. \eqref{error}) between the approximate expression for the minimum quadrature noise and the numerically computed result, as a function of the coupling coefficient and pulse duration for (a) $a=0.99$ and (b) $a=0.95$. The red circles in (a) and (b) mark the point at which the quadrature noise is at a global minimum for the given value of $a$}
\label{fig:compareDX}
\end{figure}

Generating the 3D plots in Fig. \ref{fig:minDX3d} is a relatively time-consuming process. To solve Eqs. \eqref{eq:sqamp} - \eqref{eq:thmnum} for each $\tau$ and $\kappa$, and at each time-step, we have do the integral in Eq. \eqref{eq:e3time} to obtain the pumping strength. To greatly speed-up this process we can instead use the approximate expression for $\Delta X_{\rm min} (\tau)$ given by Eq. \eqref{dxapprox}, which gives the minimum quadrature noise as a function of the peak pumping strength, $g_{\rm max}(\tau)$. The maximum value of $g_{\rm max}(\tau)$ can then be determined using the analytic expression for $g(\tau)$ given in  Eq. \eqref{pumpapprox}. The relative error between the approximate expression for the minimum quadrature noise in Eq. \eqref{dxapprox} and the numerical result is defined as, 
\begin{equation}
    \label{error}
    {\rm Error} \equiv \Bigg|\,1 - \sqrt{\frac{1+g(t_{\rm min})}{1+g_{\rm max}}}\,\Bigg|,
\end{equation}
so that when $g_{\rm max} = g(t_{\rm min})$ the error is zero. In Figs. \ref{fig:diff99} and \ref{fig:diff95} we plot the relative error as a function of $\tau$ and $\kappa$ for (a) $a=0.99$ and (b) $a=0.95$, respectively. As expected, the relative error approaches zero for long pulses. For $a=0.99$, at the optimum point (indicated by a red circle in Fig. \ref{fig:diff99}), the relative error is approximately $0.02\%$. This reinforces our assumption that $g_{\rm max}\approx g(t_{\rm min})$ when $\tau \gtrsim \tau_g$ and $a\approx1$. The relative error increases when the scattering loss increases. However, for $a=0.95$, the relative error is still only  $\approx 1\%$, indicating that the approximation can still be used confidently when $a \ge 0.95$.

\begin{figure}[htbp]
     \centering
     \includegraphics[scale=0.95]{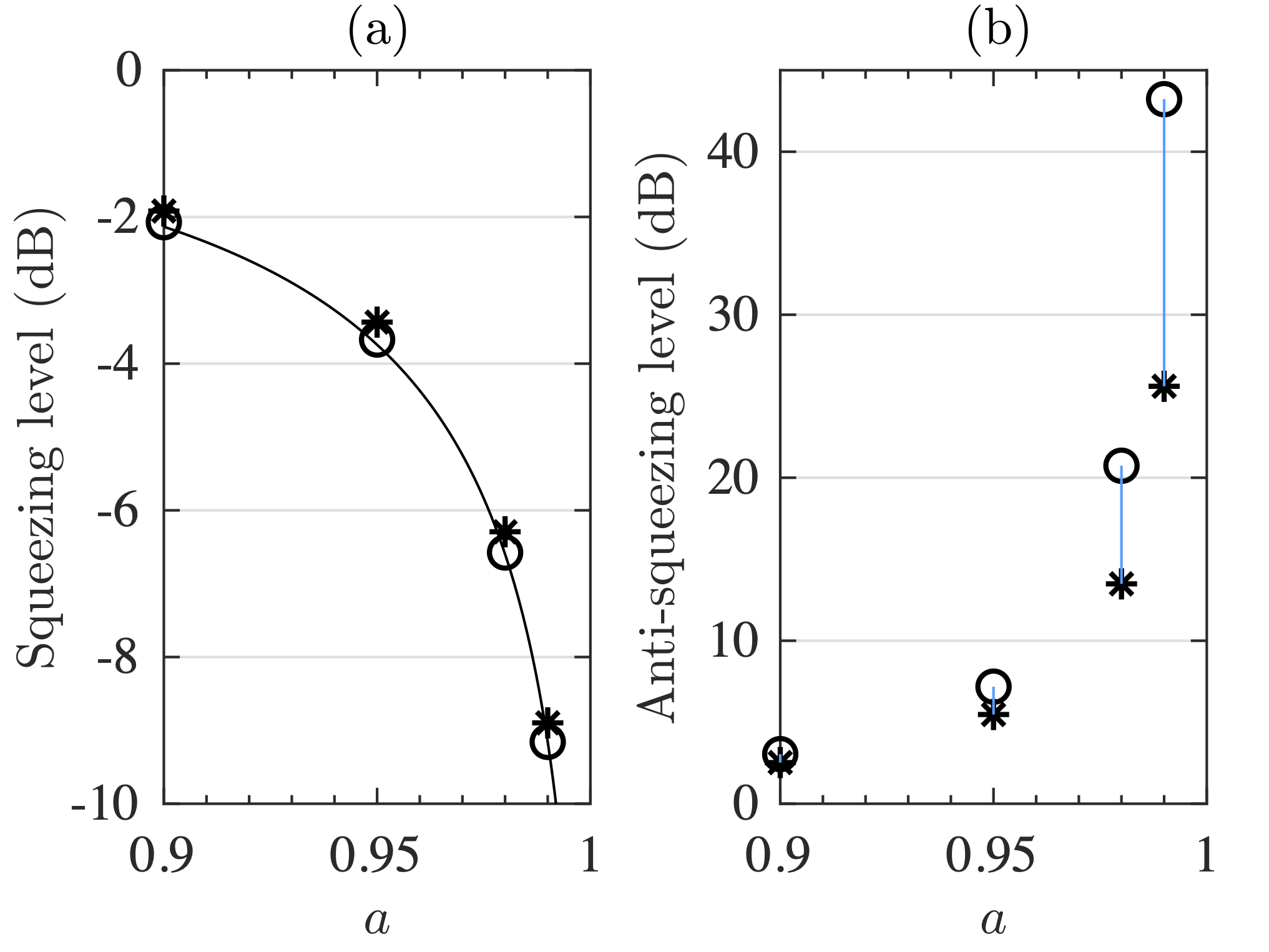}
     \caption{\small (a) The computed squeezing and (b) anti-squeezing level versus scattering loss $a$, for the optimum coupling constant $\kappa_{\rm opt}$ when $\tau = \tau_{\rm opt}$ (circles) and when $\tau = \tau_{\rm opt}/2$ (stars). The curve in (a) is our analytic expression for the squeezing, given by Eq. \eqref{dxminopt}.}
     \label{fig:dxminopt}
 \end{figure}

Letting $\sigma = \sigma_{\rm opt}$ in Eq. \eqref{dxapproxtaup}, we obtain the following approximate expression for the global minimum in the quadrature noise  $ \Delta X_{\rm opt}\equiv \Delta X_{\rm min}(\tau_{\rm opt},\sigma_{\rm opt})$ as a function of the loss parameter $a$:
\begin{equation}
    \label{dxminopt}
    \Delta X_{\rm opt} \approx \left[1+\frac{0.653g_0 }{\tilde{\Gamma}(\sigma_{\rm opt})}\sqrt{\frac{a^2-\left(1-\sqrt{3a^2+1}\right)^2}{2-\sqrt{3 a^2+1}}}\right]^{-\frac{1}{2}},
\end{equation}
where the cavity decay rate at the optimum coupling is given by,
\begin{equation}
    \label{gammaopt}
    \tilde{\Gamma}(\sigma_{\rm opt}) = -2\ln\left(-1+\sqrt{3 a^2 +1}\right).
\end{equation}
The optimum pulse duration $\tau_{\rm opt}$ is approximately given by $\tau_g(\sigma_{\rm opt})\equiv \tau_{\rm opt} $,
\begin{equation}
    \label{tauopt}
    \tilde{\tau}_{\rm opt}(a) \approx 0.342\frac{\sqrt{8\ln2}}{2-\sqrt{3a^2+1}}.
\end{equation}
The expression in Eq. \eqref{dxminopt} can be used to determine the approximate optimum squeezing level in the ring as a function of $a$.
In Fig. \ref{fig:dxminopt} (a) we compare the computed optimum squeezing level (in dB) (circle) to the value obtained with the expression in Eq. \eqref{dxminopt} (curve). As can be seen, the agreement is excellent, with a maximum relative error of $3\% $ (that is an absolute difference of $0.06 {\rm dB}$) when $a=0.9$. The globally-optimal squeezing level (for the range of $a$ considered) is approximately $-9.15 {\rm dB}$ for $a = 0.99$ and $\sigma = 0.995$. In Fig. \ref{fig:dxminopt} (b), we also show the computed anti-squeezing level (\textit{i.e.}, $\Delta Y$) (circles), when the squeezing is optimal. We see that for the global optimum in the squeezing, the anti-squeezing level is approximately $44 {\rm dB}$. Such a high level of anti-squeezing might be of concern if there is some jitter in the homodyne detection, such that one is not measuring the light at the time when it is maximally-squeezed. In the same figure, we show that by cutting the pulse duration in half (\textit{i.e.} $\tau_{\rm opt}/2$ (stars)), the anti-squeezing level reduces to approximately $26 {\rm dB}$, while the squeezing level is only modestly affected (see the stars in Fig. \ref{fig:dxminopt} (a)) (a change of less than $3\%$, or $\sim 0.3$ dB for $a=0.99$). This result is  useful for applications trying to achieve fault-tolerant quantum computing in noisy environments \cite{quantumcomputingsqueezelimit, 15dbsqueeze, Knill2005QuantumDevices}. 
 
 \subsection{Sensitivity of the minimum quadrature noise to a phase offset}
 
\begin{figure}[htbp]
     \centering
     \includegraphics[scale=1]{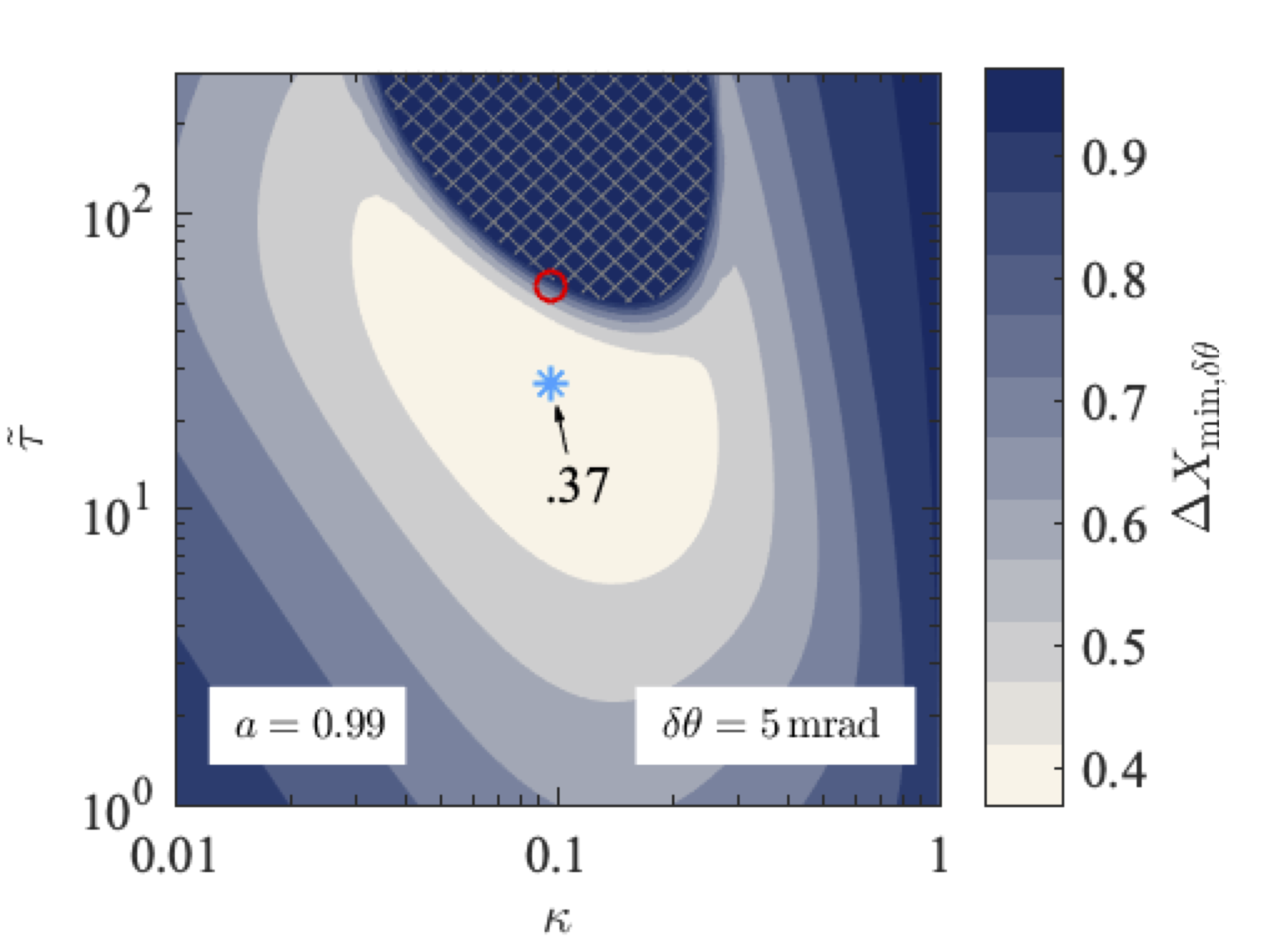}
     \caption{\small The minimum quadrature noise $\Delta X _{\delta \theta} (t_{\rm min},\tau,\kappa)$ for a phase deviation of $\delta \theta = 5\, {\rm mrad}$ as a function of the coupling constant and the pulse duration. The blue star indicates the optimal operating point, while the red circle gives the optimum point found when $\delta \theta = 0$. The hatched area indicates where the noise is greater than the vacuum noise ($\Delta X_{\delta \theta}>1$).}
     \label{fig:dxmintheta}
     
 \end{figure}

 Thus far we have assumed that the measurement of $\Delta X$ is perfect; that is, the phase of the local oscillator in a homodyne measurement is exactly matched to the phase of the squeezed light signal. We now allow for a small phase offset, $\delta \theta$, between the phase of the signal and local oscillator, and study the effect it has on the measured quadrature noise. Letting $\theta(t) = -\phi(t)/2+\delta \theta$ in the original definition for the quadrature operator in Eq. \eqref{quadopx}, the quadrature variance now is,
 \begin{eqnarray}
     \label{generalquadnoise}
     \left(\Delta X_{\delta \theta} \right)^2 &=& \left(2n_{\rm th}(t)+1\right)\times \nonumber
     \\
     &\times&\left[\cosh 2u(t) - \cos\left(2\delta\theta\right) \sinh2u(t)\right].
 \end{eqnarray}
 We interpret $\delta \theta$ as the angular deviation from the $\hat{X}$ quadrature in phase-space. If $\delta \theta = 0$ then the squeezing $\Delta X$ is measured; and if $\delta \theta = \pi/2$, then the anti-squeezing $\Delta Y$ is measured. In Fig. \ref{fig:dxmintheta}, we plot the minimum quadrature noise that is measured if the phase offset is $\delta\theta = 5\, {\rm mrad}$ and the attenuation loss in the ring is $a=0.99$. We chose this value of phase offset, because it is close to what was found in a recent experiment \cite{phasejitter}. The hatched region shows where the measured quadrature noise is greater than the vacuum noise ($\Delta X > 1$). We find that the quadrature noise has increased at the previous optimum point that we found for an offset of zero (indicated by the red circle) to $\Delta X\approx 0.8$. One can correct for the increase in noise caused by the phase offset by reducing the pulse duration to approximately $\tilde{\tau}_{\rm opt}/2\approx 26$. Doing so reduces the squeezing level to approximately $\Delta X \approx 0.37$, which is close to the optimum level for an offset of zero ($\Delta X \approx 0.35$). Note that the new optimal point (when there is phase offset) occurs for essentially the same coupling constant and only the pulse duration needs to be adjusted. Note also that there are a number of combinations of $\tau$ and $\kappa$ that achieve a squeezing level of $\Delta X <0.4$ where one could work. The results are most sensitive to a phase offset when the scattering loss is small ($a$ close to 1). For $a\le0.98$, a phase offset of $5\, {\rm m rad}$ did not significantly perturb the minimum squeezing level at $\tau_{\rm opt}$ and $\kappa_{\rm opt}$.
 
\section{Conclusion}
In this work we have studied the time-dependent squeezing process in a lossy microring resonator pumped by a Gaussian pulse. We derived approximate analytic expressions for the optimum pulse duration (Eq. \eqref{taup_text}) and optimum ring-channel coupling constant (Eq. \eqref{eq:sigmaopt}) for a fixed pump energy. Using these optimal parameters, we derived an analytic expression for the maximum squeezing level achievable for a ring with a given loss $a$ (Eq. \eqref{dxminopt}). We found that for the chosen pump energy of $0.188\,$pJ and a scattering loss of $a=0.99$, the optimal coupling constant and pulse duration are $\sigma_{\rm opt} = 0.995$ and $\tau_{\rm opt} = 56 T_R$, while for a scattering loss of $a=0.95$ we find optimal values of $\sigma_{\rm opt} = 0.974$ and $\tau_{\rm opt} = 13T_R$. Under these optimal conditions, we demonstrated a maximum squeezing level of $-9.15{\rm dB}$ and $-3.67 {\rm dB}$ for $a=0.99$ and $a=0.95$, respectively. Furthermore, we demonstrated that by reducing the pulse duration at optimal coupling, the anti-squeezing level can be drastically reduced, while the squeezing level is only modestly affected. Moreover, we showed that our model shows how one can reduce the impact of homodyning phase noise on the squeezing simply by reducing the pump pulse duration from the nominally optimal value. We believe that the analytic expressions that we have developed for this system will help researchers looking to optimize the design of ring resonator systems for the generation of squeezed light. 

\label{sec:conclusions}
\section*{Acknowledgements}
This work was supported by Queen’s University and the Natural Sciences and Engineering Research Council of Canada (NSERC). The authors would also like to thank Hossein Seifoory for many fruitful discussions.


\bibliography{apssamp}


\appendix
\section{Derivation of the time-dependent pump pulse in the ring}
\label{ringpulsederiv}
In this section we  derive an approximate expression for the pump field in the ring, $E_3\left(\tilde{t}\,\right)$. We start with Eq. \eqref{eq:e3time}. To simplify this, we define 
\begin{equation}
\label{chiomega}
    \chi(\Omega) \equiv  \frac{1}{\exp(-i\Omega)-\sigma a}.
\end{equation}
 We expand this in a Taylor series about $\Omega=0$,
 \begin{eqnarray}
 \label{taylor}
 \chi(\Omega) = \sum_{n=0}^{\infty}\frac{\Omega^n\chi^{(n)}(0)}{n!},
 \end{eqnarray}
 where 
 \begin{equation}
 \label{taylorderiv}
     \chi^{(n)}(0)\equiv \frac{d^n \chi(\Omega)}{d\Omega^n}\bigg|_{\Omega = 0}
 \end{equation}
 is the $n^{th}$ order derivative of $\chi$ evaluated at $\Omega = 0$.
In the high squeezing limit $(1-\sigma a)\ll 1$ it can be shown that for $n\ge 2$, the $n^{ th}$ and $(n-2)^{th}$ derivatives are related by,
 \begin{equation}
 \label{derivsrelation}
     \frac{\chi^{(n)}(0)}{n!} = -\frac{1}{\epsilon^2}\frac{\chi^{(n-2)}(0)}{(n-2)!},
 \end{equation}
where $\epsilon \equiv 1- \sigma a$. Using 
Eq. \eqref{derivsrelation} in Eq. \eqref{taylor} and after simplifying we find that we can write $\chi(\Omega)$ as
 \begin{eqnarray}
\label{chiomegaapprox}
  \chi(\Omega)&=& \frac{1}{\epsilon}\left(\frac{1+ i\Omega/\epsilon }{1+\Omega^2/\epsilon^2}\right).
 \end{eqnarray}
 The modulus-squared of this is a good approximation to the buildup factor around the peak at $\Omega =0$. Now we define the integral in Eq. \eqref{eq:e3time} as $A\left(\tilde{t}\,\right)$. It is given by, 
 \begin{eqnarray}
     \label{integralofnote1}
     A\left(\tilde{t}\,\right)&=&\exp\left(\frac{-2\ln(2)\tilde{t}^2}{\tilde{\tau}^2}\right)\int_{-\infty}^{\infty} d\Omega\chi(\Omega)\times  \nonumber
     \\
     &\times&\exp\left[-\left(\frac{\Omega \tilde{\tau}}{\sqrt{8\ln2}}+i\frac{\sqrt{8\ln{2}}\,\tilde{t}}{2\tilde{\tau}}\right)^2\right],
 \end{eqnarray}
 where we have completed the square in the argument of the exponential in Eq. \eqref{eq:e3time} to get this form.
 Using Eq. \eqref{chiomegaapprox} in Eq. \eqref{integralofnote1}, we obtain
 \begin{eqnarray}
     \label{integralofnote2}
     A\left(\tilde{t}\,\right)&=&\frac{1}{\epsilon}\exp\left(\frac{-2\ln(2)\tilde{t}^2}{\tilde{\tau}^2}\right)\int_{-\infty}^{\infty} d\Omega\frac{1+ i\Omega/\epsilon }{1+\Omega^2/\epsilon^2}\times  \nonumber
     \\
     &\times&\exp\left[-\left(\frac{\Omega \tilde{\tau}}{\sqrt{8\ln2}}+i\frac{\sqrt{8\ln{2}}\,\tilde{t}}{2\tilde{\tau}}\right)^2\right].
 \end{eqnarray}
Now we make the following substitutions in Eq. \eqref{integralofnote2}: $y=\Omega/\epsilon$, $s=2\ln(2)/(\epsilon^2\tilde{\tau}^2)$, and $x = -i\tilde{t}4\ln(2)/(\epsilon\tilde{\tau}^2)$ . Doing this we obtain,
\begin{eqnarray}
    \label{integralofnote3}
   A\left(\tilde{t}\,\right)&=&\exp\left(\frac{-2\ln(2)\tilde{t}^2}{\tilde{\tau}^2}\right)\times
    \\
    &\times&\int_{-\infty}^{\infty}dy\left[\frac{{\rm e}^{-(x-y)^2/(4s)}}{1+y^2}+ i\frac{y{\rm e}^{-(x-y)^2/(4s)}}{1+y^2}\right] .\nonumber
\end{eqnarray}
The integrals in Eq. \eqref{integralofnote3} can be expressed in terms of Voigt functions $U(x,s)$ and $V(x,s)$ \cite{NIST:DLMF}: 
\begin{eqnarray}
\label{Uvoigt}
    U(x,s) = \frac{1}{\sqrt{4\pi s}}\int_{-\infty}^{\infty}dy\frac{{\rm e}^{-(x-y)^2/(4s)}}{1+y^2},
\end{eqnarray}
and
\begin{eqnarray}
\label{Vvoigt}
    V(x,s) = \frac{1}{\sqrt{4\pi s}}\int_{-\infty}^{\infty}dy\frac{y{\rm e}^{-(x-y)^2/(4s)}}{1+y^2}.
\end{eqnarray}
It can be shown that
\begin{equation}
    \label{VoigtIdentity}
    U(x,s)+iV(x,s) = \sqrt{\frac{\pi}{4s}}{\rm e}^{z^2}{\rm erfc}\,z
\end{equation}
with $z = (1-ix)/(2\sqrt{s})$. The Eqs. \eqref{Uvoigt} - \eqref{VoigtIdentity} allow us to write Eq. \eqref{integralofnote3} as,
\begin{eqnarray}
    \label{integralofnotefinal}
       A\left(\tilde{t}\,\right) = \exp\left(\frac{-2\ln(2)\tilde{t}^2}{\tilde{\tau}^2}\right)\pi{\rm e}^{z\left(\tilde{t}\,\right)^2}{\rm erfc}\,z\left(\tilde{t}\,\right).
\end{eqnarray}
 Transforming back to our original variables $\tilde{\tau}$ and $\tilde{t}$ we obtain $z\left(\tilde{t}\,\right)  = (1-\sigma a)\tilde{\tau}/{\sqrt{8\ln(2)}}-\sqrt{8\ln(2)}\tilde{t}/(2\tilde{\tau})$. Replacing the integral in Eq. \eqref{eq:e3time} with the expression in Eq. \eqref{integralofnotefinal} gives Eq. \eqref{e3timeapprox} in the text.
\section{Derivation of $\tau_g$}
\label{Doftaug}
In this section we derive an approximate expression, Eq. \eqref{taup_text} for the pulse duration $\tau_g$ that gives the peak in the pumping strength (see Fig. \ref{fig:e3time}). In order to do this, we first hold $\tau$ constant and then find the time $t_{peak}$ when the pump is at its peak value. Then we determine the pulse duration that causes the greatest peak value. We solve the following two equations simultaneously;
\begin{eqnarray}
    \label{gdifft}
    \frac{\partial g(t,\tau)}{\partial t}\bigg|_{t=t_{peak}} = 0,
\end{eqnarray}
and,
\begin{eqnarray}
\label{gdifftau}
    \frac{\partial g(t_{peak},\tau)}{\partial \tau}\bigg|_{\tau = \tau_g} = 0.
\end{eqnarray}
Re-writing Eq. \eqref{e3timeapprox} in terms of $z(t)$ alone and ignoring the factors that do not depend on $t$ or $\tau$, we find
\begin{equation}
    \label{gz}
    g(\tilde{t}) \propto \sqrt{\tilde{\tau}}\exp\left(-\frac{\epsilon^2\tilde{\tau}^2}{8\ln2}+\frac{2\epsilon\tilde{\tau}z(\tilde{t})}{\sqrt{8\ln2}}\right){\rm erfc}\,z(\tilde{t}),
\end{equation}
where $\epsilon\equiv 1-\sigma a$, $\tilde{\tau}=\tau/T_R$, and $\tilde{t}=t/T_R$. Also $T_R\partial/\partial t = \partial/\partial \tilde{t}$ and $\partial/\partial\tilde{t} = -(\sqrt{2\ln2}/\tilde{\tau})\partial/\partial z$. Using Eq. \eqref{gz} in Eq. \eqref{gdifft} and switching the derivatives to $z$, we obtain the following implicit equation for $z(\tilde{t}_{peak})$;
\begin{equation}
\label{impliciteq}
    {\rm e}^{z_{peak}^2}{\rm erfc}\,z_{peak} = \frac{1}{\sqrt{\pi}}\frac{\sqrt{8\ln2}}{\epsilon \tilde{\tau}},
\end{equation}
where $z_{peak}\equiv z(\tilde{t}_{peak})$. Now, using Eq. \eqref{gz} in Eq. \eqref{gdifftau} and noting that $T_R\partial/\partial \tau = \partial/\partial \tilde{\tau}$, we obtain
\begin{eqnarray}
\label{maxzt}
0&=&\frac{1}{2\tilde{\tau}_g}-\frac{\tilde{\tau}_g\epsilon^2}{4\ln2}+\frac{2\epsilon z_{peak}}{\sqrt{8\ln 2}}+\nonumber
\\
&+&\left(\frac{2\tilde{\tau}_g\epsilon}{\sqrt{8\ln 2}} - \frac{2}{\sqrt{\pi}}\left[ {\rm e}^{z_{peak}^2}{\rm erfc}\,z_{peak}\right]^{-1}\right)\frac{\partial z_{peak}}{\partial \tilde{\tau}}\bigg|_{\tilde{\tau}_g}, \nonumber
\\
0&=&\frac{1}{2\tilde{\tau}_g}-\frac{\tilde{\tau}_g\epsilon^2}{4\ln2}+\frac{2\epsilon z_{peak}}{\sqrt{8\ln 2}},
\end{eqnarray}
where the second equation is obtained from the first by using Eq. \eqref{impliciteq}. Solving Eq. \eqref{maxzt} for $z_{peak}$ gives,
\begin{equation}
\label{maxztfinal}
    z(\tilde{t}_{peak}) = \frac{\epsilon \tilde{\tau}_g}{\sqrt{8\ln2}} - \frac{\sqrt{8\ln 2}}{4\epsilon\tilde{\tau}_g}.
\end{equation}
Transforming Eq. \eqref{maxztfinal} back to time $t$ and using Eq. \eqref{ztransformation} we find that
\begin{equation}
    \label{maxt}
    \tilde{t}_{peak}(\tilde{\tau}_g) = \frac{1}{2(1-\sigma a)}
\end{equation}
is the time when $g(\tau_g)$ is at its peak value. The time $\tilde{t}_{peak}$ is the inverse of the decay rate $1/\tilde{\Gamma}$ in the low-loss limit $(1-\sigma a)\ll 1$. We can determine $\tilde{\tau}_g$ by using Eq. \eqref{maxztfinal} in Eq. \eqref{impliciteq}. Doing this gives the following transcendental equation:
\begin{equation}
\label{numericaleq}
    \exp\left(x-\frac{1}{4x}\right)^2{\rm erfc}\left(x-\frac{1}{4x}\right)=\frac{1}{\sqrt{\pi}x},
\end{equation}
where $x\equiv \epsilon \tilde{\tau}_g/(8\ln 2)^{1/2}$. We have numerically determined that the solution of Eq. \eqref{numericaleq} is $x\approx 0.34189$. Thus, $\tilde{\tau}_g$ is approximately given by
\begin{equation}
    \label{taup}
    \tilde{\tau}_g \approx0.342\frac{\sqrt{8\ln2}}{1-\sigma a},
\end{equation}
which is the expression given in Eq. \eqref{taup_text}.
\end{document}